\documentclass{article}
\usepackage{arxiv}
\usepackage[utf8]{inputenc} 
\usepackage[T1]{fontenc}    
\usepackage{hyperref}       
\usepackage{url}            
\usepackage[most]{tcolorbox}
\usepackage{booktabs}       
\usepackage{amsfonts}       
\usepackage{amsmath}
\usepackage{nicefrac}       
\usepackage{microtype}      
\usepackage{bm}             
\usepackage{lipsum}
\usepackage{enumitem}
\usepackage{tcolorbox}
\usepackage{graphicx}
\usepackage{booktabs}
\usepackage{multirow}
\usepackage{appendix}
\usepackage{threeparttable}
\newcounter{algocounter}
\graphicspath{ {./images/} }
\usepackage{epstopdf}
\ifpdf
  \DeclareGraphicsExtensions{.eps,.pdf,.png,.jpg}
\else
  \DeclareGraphicsExtensions{.eps}
\fi

\usepackage{algorithm}
\usepackage{algpseudocode}
\usepackage{threeparttable}
\usepackage{caption}
\usepackage{subcaption}
\usepackage{mathtools}
\usepackage{amssymb}
\usepackage{multirow}
\usepackage{siunitx}

\setlist[enumerate]{leftmargin=.5in}
\setlist[itemize]{leftmargin=.5in}

\newenvironment{codefont}{\fontfamily{lmtt}\selectfont}{\par}
\DeclareTextFontCommand{\codetext}{\codefont}

\title{SuSIE: Subset Simulation with Intrepid Exploration}

\author{
 Adwait Sharma \\
  Hopkins Extreme Materials Institute\\
  Johns Hopkins University\\
  Baltimore, MD 21218 \\
  \texttt{ashar125@jhu.edu} \\
   \And
 Promit Chakroborty \\
  Department of Civil and Environmental Engineering\\
  Vanderbilt University\\
  Nashville, TN 37203 \\
  \texttt{promit.chakroborty@vanderbilt.edu} \\
   \And
 Michael D. Shields \\
  Department of Civil and Systems Engineering\\
  Johns Hopkins University\\
  Baltimore, MD 21218 \\
  \texttt{michael.shields@jhu.edu} \\
}

\begin{document}

\newtheorem{proof}{Proof}
\newtheorem{theorem}{Theorem}
\newtheorem{lemma}{Lemma}
\newtheorem{corollary}{Corollary}
\newtheorem{proposition}{Proposition}
\newtheorem{definition}{Definition}
\newtheorem{remark}{Remark}
\newtheorem{stipulation}{Stipulation}

\maketitle

\begin{abstract}

This work explores the challenges associated with the subset simulation framework - a well-established algorithm for estimating structural reliability - in settings involving multiple, possibly disconnected regions of failure, or involving discontinuous or sharply changing performance functions. We demonstrate that these drawbacks of subset simulation stem from the limitations of the Markov chain Monte Carlo (MCMC) sampler employed within the method, not from the subset simulation framework itself. Traditional random-walk Metropolis algorithms, which are known to struggle severely with sampling from multimodal distributions, are conventionally applied within subset simulation, which leads to inaccurate failure probability estimates in such cases. In this work, we instead utilize a modified version of the Intrepid MCMC sampler, which has recently been shown to be more effective than vanilla random-walk Metropolis algorithms in sampling from multimodal probability distributions. The subset simulation method with the proposed Intrepid sampler is demonstrated to address these complicating features. Several illustrative examples are considered, ranging from 2 to 1003 dimensions and exhibiting multiple failure regions or highly nonlinear performance functions, including both analytical problems and structural engineering applications.

\end{abstract}


\section{Introduction}
\label{section:introduction}

We consider the problem of computational structural reliability modeling, which involves estimating the probability of failure, given as follows: 
\begin{equation}
\label{eqn:reliability_definition}
    P_F = \mathbb{P}(g(\bm{X}) \leq 0) = \int_{-\infty}^{\infty} I \left[ g(\bm{x}) \leq 0 \right] f_{\bm{X}}(\bm{x}) \, d\bm{x}.     
\end{equation}
Here, uncertainties associated with the structural system and applied loads are modeled as an $n_d$-dimensional random vector $\bm{X}$ having a joint probability density function (PDF) $f_{\bm{X}}(\bm{x})$. The performance function $g(\bm{x})$ is defined such that $g(\bm{x}) \leq 0$ indicates that the structure has failed; consequently, the indicator function $I[\cdot]$ takes value $ 1 $ when $g(\bm{x}) \leq 0$, i.e., failure occurs, and $ 0 $ otherwise. The random vector $\bm{X}$ is often transformed into the standard normal random vector $\bm{U}$ via Rosenblatt's~\cite{rosenblatt1952remarks} or Nataf’s~\cite{nataf1962determination} transformation, and the reliability integral is rewritten in the standard normal space as:
\begin{equation}
\label{eqn:reliability_definition_gaussian}
    P_F = \int_{-\infty}^{\infty} I [ G(\bm{u}) \leq 0 ] \phi(\bm{u}) \, d\bm{u}
\end{equation}
Here, $\phi(\bm{u})$ is the $n_d$-dimensional standard normal PDF and $G(\bm{u})$ is the performance function defined in the standard normal space. 

Early techniques for estimating failure probabilities, such as the first and second-order reliability methods (FORM and SORM)~\cite{hasofer1974exact,breitung1984asymptotic,melchers2018structural} lead to an approximate estimate of $P_F$ based on the Taylor series approximation of the performance function around design points - points on the limit surface (i.e., $G(\bm{u}) = 0$) with the highest PDF value - obtained as solutions to a constrained optimization problem. Even though SORM, in particular, has been shown to be asymptotically exact as the Hasofer-Lind reliability index (which is defined as the Euclidean distance of the design point from the origin) tends to infinity, quantifying the error associated with the failure probability estimate produced by these methods is not straightforward. Additionally, the estimate produced by these methods cannot be improved even if the analyst is willing to expend more computational effort~\cite{rackwitz2001reliability,hohenbichler1988improvement}. With the emergence of increasingly powerful computational tools, it has become more desirable to adopt methods whose estimation accuracy can be improved when more computational resources are invested. 

The direct Monte Carlo (MC) method achieves this goal by estimating the failure probability statistically, recasting the failure probability integral in Eq.~\eqref{eqn:reliability_definition_gaussian} as the finite sample average
\begin{equation}
    P_F \approx \dfrac{1}{N} \sum_{i=1}^N I [ G(\bm{u}^{(i)}) \leq 0 ]
\end{equation}
This produces an estimator that (a) converges with probability $1$ to the true failure probability as $N \rightarrow \infty$, (b) has a sampling distribution characterized by the Central Limit Theorem, and (c) has statistical properties that are independent of $n_d$. While these features are highly desirable in an estimator, it is well known that the direct MC estimator demands a large number of performance function evaluations when estimating small failure probabilities (typically of the order $10^{-4}$ or lower), which frequently arise in engineering problems. This is especially infeasible because in engineering applications, each call to the performance function often involves running a computationally intensive simulation. 

Given this difficulty, the reliability literature broadly presents two main approaches to estimate failure probabilities using MC with reasonable computational effort.
The first approach, which is the focus of the present study, involves constructing an estimator that achieves the same sampling variance with fewer samples compared to MC, and therefore fewer performance function calls. These methods are termed variance reduction techniques~\cite{au2014engineering,tabandeh2022review,song2023monte}.
The second approach involves constructing a computationally cheaper surrogate model of the performance function or the limit surface. The original performance function (or limit surface) is then replaced by this surrogate model on which MC methods can be performed with reduced computational effort~\cite{moustapha2022active,afshari2023deep}.


\subsection{Exploration vs. exploitation in reliability estimation}
\label{subsection:challenges}

Variance reduction techniques typically rely on identifying important regions of failure, either through an adaptive or a preliminary exploration, and leveraging this information to estimate failure probabilities with a small number of samples. The idea here is to avoid sampling from regions that do not contribute to the failure probability, and to invest the limited computational effort in sampling in regions where failures occur. Surrogate modeling techniques, similarly, aim to locate important regions of failure and accurately and efficiently approximate the performance function in these regions. The intuition underlying both strategies, therefore, is to explore the space to identify the the highest probability regions where failure occurs, and concentrate effort in this region.
In fact, even FORM and SORM rely on locating the design point, the point along the limit surface with maximum probability.

The sampling schemes underlying these methods exploit trends in the performance function to identify important regions in the input domain, i.e., regions around the limit surface that also have a significant probability of occurrence.  However, this focus on exploitation
renders these methods vulnerable to insufficient exploration of the space. Insufficient exploration becomes a major challenge when the performance function has: (a) several regions of failure, which may be disconnected or contribute unequally to failure probability, and/or (b) discontinuities or sharp changes in performance function values~\cite{breitung2019geometry,breitung2021sorm,rashki2021sesc,sharma2023modified}. When these geometric features are present, the sampler may be misled into focusing only on a subset of the important regions of input space, getting stuck in the identified regions and failing to discover others.
In contrast, the direct MC method possesses no mechanism for exploitation.
Instead, the entire space is thoroughly explored, 
make it immune to the aforementioned geometric difficulties. Unfortunately, this exploration comes at the cost of an enormous number of samples, including many samples in regions that do not meaningfully contribute to the failure probability. 
Therefore, 
the success of any reliability estimation technique hinges on its ability to balance exploring the space to find all important regions of failure and exploiting the known structure of the performance function to focus on important failure regions.

\subsection{Subset simulation and its limitations}
\label{section:subset_simulation}

Originally proposed by Au and Beck~\cite{au2001estimation}, subset simulation (SuS) has become one of the most effective and extensively studied methods for
structural reliability~\cite{ching2005reliability,katafygiotis2007application,zuev2012bayesian,cui2019implementation}. 
SuS operates by defining a decreasing sequence of intermediate failure thresholds $b_1>b_2>\cdots>b_{n_f}=0$, and corresponding intermediate failure regions $F_i=\{\bm{u} \in \mathbb{R}^{n_d}: G(\bm{u})\leq b_i\},i=1,2,...,n_{f}$, where $F_1\supset F_2 \supset \cdots \supset F_{n_f}$, and $F_{n_{f}}=F=\{\bm{u} \in \mathbb{R}^{n_d}: G(\bm{u})\leq 0\}$ denotes the actual failure region. 
The probability of failure $P_F=\mathbb{P}(F)$ is then rewritten as:
\begin{equation}
\label{eqn:subsim_product}
P_F=\mathbb{P}(F)=\mathbb{P}(F_1)\mathbb{P}(F_2\mid F_1)\cdots\mathbb{P}(F_{n_f}\mid F_{n_f-1}),
\end{equation}
where the conditional probability $\mathbb{P}(F_{l+1}\mid F_l)$, $l=1,2,...,n_{f}-1$, also called the intermediate failure probability for subset $l+1$, is the probability that a point in subset $F_l$ also lies in subset $F_{l+1}$. When the failure thresholds, $b_i$, are specified a priori, these conditional probabilities are estimated using MC from samples drawn from conditional standard normal distributions of the form
\begin{equation}
\label{eqn:sus_subset_conditional_distributions}
    \phi(\bm{u}\mid{F_l}) \propto I [ G(\bm{u}) \leq b_l ] \phi(\bm{u}).
\end{equation}
However, in practice, the conditional probabilities are assigned a predetermined value, i.e., $\mathbb{P}(F_{l+1}\mid F_{l})=p_0$ for $l=1,2,\cdots,n_f-1$ (typically $p_0=0.1 \text{ to } 0.3$~\cite{zuev2012bayesian}), and the associated intermediate threshold values $b_{l+1}$ are determined adaptively
by drawing samples using Markov Chain Monte Carlo (MCMC) according to $\phi(\bm{u}\mid{F_l})$ and computing the $p_0$ percentile threshold. 

The conditional PDFs $\phi(\bm{u}\mid{F_l})$ 
are discontinuous along the intermediate limit surface $ G(\bm{u})= b_l $. As a result, the geometries of these distributions may be highly complex, with multiple disconnected and distant modes of unknown probability.
Traditionally, samples from the previous conditional distribution $\phi(\bm{u}\mid{F_{l-1}})$ are used as seeds for standard MCMC on $\phi(\bm{u}\mid{F_l})$, with the assumption that these seeds sufficiently cover the support of $\phi(\bm{u}\mid{F_l})$ to allow adequate sampling with short chain lengths.
However, when the performance function is discontinuous or changes rapidly, standard MCMC approaches may fail to sample from the complex conditional distribution for several reasons. Certain regions of importance may lack seed samples altogether. When these regions are disconnected from other modes and separated by distances larger than the characteristic length scale of the proposal distribution, they may never be practically discoverable.  Even when these regions are connected to modes with representative seed samples, they may be connected only by very low probability paths that require very long chains to discover. This issue has long been known and was, in fact, identified in the original study by Au and Beck~\cite{au2001estimation}.
As a result, probability of failure estimates from SuS can have large variability, since sometimes all modes are found and sometimes they are not. It is now well-documented in the literature that, in the presence of such features, SuS often performs poorly~\cite{breitung2019geometry,rashki2021sesc,sharma2023modified,kinnear2025niching}.

In fact, recent studies have criticized SuS~\cite{breitung2019geometry,breitung2021sorm} because its standard implementation is unable to effectively handle these challenges. However, as discussed above, these challenges are not fundamental to the SuS framework but rather stem from the difficulties in appropriately sampling from the complex conditional distributions. 
The SuS framework of expressing the true failure probability as a product of conditional failure probabilities is theoretically sound and rigorously developed. The problem inherently lies in developing MCMC schemes capable of sampling from complex distributions.
In fact, development of advanced MCMC sampling schemes
for SuS, including variants of the Metropolis-Hastings (MH) algorithm~\cite{santoso2011modified,zuev2011modified,papaioannou2015mcmc}, Hamiltonian Monte Carlo~\cite{wang2019hamiltonian,chen2022riemannian,thaler2024reliability}, and ensemble samplers~\cite{shields2021subset}, has been of significant interest, and has demonstrated improved performance for many problems.
Likewise, a scheme suited to sampling multimodal distributions will be able to address many of the issues raised in~\cite{breitung2019geometry,breitung2021sorm}.

\subsection{Contributions of this study}
\label{subsection:contributions}

It is well known that the traditional random-walk Metropolis (RWM) sampler struggles severely with sampling from multimodal or complexly-shaped distributions~\cite{Geyer1991,neal2011mcmc}. This has spawned a great deal of research into MCMC algorithms 
for this class of problems~\cite{pompe2020,latuszynski2026mcmc,chakroborty2026intrepid}. However, most of these methods lose the attractive simplicity of RWM formulations through the introduction of multiple difficult-to-tune parameters, the use of complicated proposals that are challenging to construct, or intermediate targets that result in extraneous computations~\cite{Geyer1991,ahn2013distributed,gabrie2022adaptive}. Moreover, development of MCMC samplers that overcome these challenges specifically in the context of structural reliability 
has received limited attention in the literature. These challenges are further exacerbated when the problem is high dimensional. The replica exchange MCMC sampler proposed by Sharma and Manohar~\cite{sharma2023modified}, for example, has been shown to alleviate these geometric difficulties in lower dimensions. However, existing MCMC samplers used in SuS do not address these challenges adequately in large dimensions.

The recently introduced Intrepid MCMC~\cite{chakroborty2026intrepid} algorithm, which injects exploratory steps within a RWM framework, has demonstrated improved robustness when sampling from multi-modal distributions while retaining the simplicity of RWM algorithms. This combination of simplicity and exploratory robustness is attractive for use in SuS.
However, the authors in~\cite{chakroborty2026intrepid} show that the acceptance rate of the exploratory steps diminishes in high dimensions, reducing its  exploration capacity. 
In this work, we propose a cyclic component-wise Intrepid MCMC for SuS, which has been specifically designed to improve exploration in high dimensions while drawing samples from conditional standard normal PDFs. This leads to the following key contributions:
\begin{enumerate}
    \item Intrepid MCMC, a novel variant of the RWM algorithm that is significantly more robust at sampling from multi-modal distributions than standard RWM, is integrated into SuS.
    \item A cyclic component-wise implementation of the Intrepid MCMC algorithm is proposed to improve exploration in high dimensions.
    \item Improvement in SuS performance with Intrepid MCMC is showcased for problems involving multiple failure regions and complex performance function geometries.     
\end{enumerate}
These contributions demonstrate that the limitations of subset simulation observed in~\cite{breitung2019geometry,breitung2021sorm} are not rooted in the SuS framework itself, but rather in poorly performing MCMC sampling algorithms. While this has been noted in the literature~\cite{breitung2021sorm,li2025relaxed}, to the best of our knowledge, no methods currently exist that deal with these difficulties within the original SuS framework solely by replacing the MCMC sampler. Instead, existing methods to address these challenges either modify the algorithmic framework of SuS~\cite{rashki2021sesc,kinnear2025niching,li2025relaxed} or require precursor runs to tune auxiliary parameters~\cite{sharma2023modified}.

\section{Cyclic component-wise Intrepid MCMC for subset simulation}
\label{section: componentwise}

\subsection{A brief summary of Intrepid MCMC}

Intrepid MCMC, recently proposed by Chakroborty and Shields~\cite{chakroborty2026intrepid}, is a variant of the RWM algorithm aimed at improving sampling robustness and convergence rate for complex, multimodal target distributions that can be expressed in the form, 
\begin{equation}
\label{eqn:intrepid_target_pdf}
    \pi(\bm{x}) \propto T(\bm{x})p(\bm{x})
\end{equation}
where $p(\bm{x})$, called the "parent distribution," is a known simple distribution (e.g.~Gaussian) and $T(\bm{x})$ is a transformation function that encodes the physical or simulation process.
We consider the context of subset simulation in the standard normal space, where we have $p(\bm{x}) =\phi(\bm{u})$ and $T(\bm{x}) = I [ G(\bm{u}) \leq b_l ]$, which results in the conditional PDF of interest $ \phi(\bm{u}\mid{F_l}) $ defined in Eq.~\eqref{eqn:sus_subset_conditional_distributions}. Depending on the geometric nature of the indicator function $I [ G(\bm{u}) \leq b_l ]$, $ \phi(\bm{u}\mid{F_l}) $ could have multiple modes with no prior knowledge of their locations or their relative importance.

The Intrepid sampler draws from $\phi(\bm{u}\mid{F_l})$ by employing a mixture distribution~\cite{tierney1994markov} of two Markov transition PDFs:
\begin{enumerate}
    \item A local transition PDF, $p_{\text{local}}(\cdot\mid \cdot)$ is focused on generating samples from the neighborhood of the current Markov chain state. This transition corresponds to that of the standard Cartesian component-wise Metropolis algorithm proposed by Au and Beck~\cite{au2001estimation} and therefore does not warrant further discussion.
    \item A global transition, $p_{\text{global}}(\cdot\mid \cdot)$ explores the space in search of new, potentially undetected modes. This is achieved by generating proposal states which are perturbed along constant probability contours of the parent distribution $\phi(\bm{u})$. 
For $\phi(\bm{u})$, these contours are hyperspheres centered at the origin. 
The Intrepid sampler, therefore, constructs a global transition PDF that executes MH steps on the hypersphere, rather than in Cartesian coordinates. By proposing MH transitions in hyperspherical coordinates, $p_{\text{global}}(\cdot\mid \cdot)$ proposes transitions between regions of approximately equal importance that may lie far away from each other (in terms of, say, Euclidean distance), which would be virtually inaccessible to each other through local transitions only in the Cartesian components.
\end{enumerate}

The Intrepid Markov transition PDF is defined as the mixture of $p_{\text{local}}(\cdot\mid \cdot)$ and $p_{\text{global}}(\cdot\mid \cdot)$ given by
\begin{equation}
\label{eq:original_intrepid_mtd}
p_{\text{Intrepid}}\left(\bm{u}^{(k+1)}\mid \bm{u}^{(k)}\right)=(1-\beta) p_{\text{local}}\left(\bm{u}^{(k+1)}\mid \bm{u}^{(k)}\right)+ \beta p_{\text{global}}\left(\bm{u}^{(k+1)}\mid \bm{u}^{(k)}\right),
\end{equation}
where $\beta$ is a hyperparameter that defines the probability of making a global transition. 

In the standard normal space, given a state $\bm{u}$ with Cartesian coordinates $\bm{u}=\left(u_{1},u_{2},\cdots,u_{n_d}\right)$ and hyperspherical coordinates $\left(r,\theta_{1},\cdots,\theta_{n_d-1}\right)$, the component-wise local proposal PDFs may be written as
\begin{equation}
    \label{eq:local_proposal}
    q_{\text{local},i}(\tilde u_{i}\mid u_{i})
    =
    \frac{1}{\sqrt{2\pi}}
    \exp\left[
    -\frac{\left(\tilde u_{i}-u_{i}\right)^2}{2}
    \right], \text{for } i=1,2,\cdots,n_d.
\end{equation}
The global proposal takes the form~\cite{chakroborty2026intrepid}
\begin{equation}
    \label{eq:global_full_proposal}
    q_\text{global} (\Tilde{\bm{u}} | \bm{u}) = \frac{q_{\gamma} (\gamma) \left[ \prod_{i=1}^{n_d-1} q_{\Theta_{i}} \left( \phi_{i} | \theta_{i} \right) \right]}{\left[ r \Tilde r ^{n_d-1} \right] \left[ \prod_{i=1}^{n_d-1} \sin^{\left( n_d-i-1 \right)} \left( \Tilde \theta_{i} \right) \right]}
\end{equation}
where $\gamma = \Tilde r/r$ is the radial perturbation factor, $\phi_{i}=\Tilde \theta_{i} - \theta_{i},i=1,2,\cdots,n_d-1$ are the angular perturbations, and the hyperspherical component-wise proposal PDFs may be defined as follows
\begin{equation}
        \begin{aligned}
            \label{eqn:angular_proposal_symmetric_choices}
                q_{\gamma} (\gamma) &= \text{Uniform}(1/\gamma_0,\gamma_0) \\
                q_{\Theta_{i}} \left( \phi | \theta_{i} \right) &= \text{Uniform} \left( - \theta_{i} , \pi - \theta_{i} \right) \text{for } i=1,2,\cdots,n_d-2 \\
                q_{ \Theta_{n_d-1} } \left( \phi | \theta_{n_d-1} \right) &= \text{Uniform} \left( - \theta_{n_d-1} , 2 \pi - \theta_{n_d-1} \right)
        \end{aligned}
\end{equation}
The recommended value of $\gamma_0$ is 2~\cite{chakroborty2026intrepid}. The transition PDF is then constructed by combining these proposal densities with a standard MH acceptance PDF, $\alpha(\Tilde{u},u)$ such that 

\begin{align}
        \label{eqn:local_component_tpdf}
        p_{\text{local},i}(\tilde{u}_i\mid u_i) &= \alpha(\Tilde{u}_i,u_i)q_{\text{local},i}(\tilde u_{i}\mid u_{i}) + \left[1-\int{\alpha(v_i,u_i)q_{\text{local},i}(v_{i}\mid u_{i})}\,dv_i \right]\delta(\tilde u_{i}-u_{i})\\
        \label{eqn:local_complete_tpdf}
        p_{\text{local}}(\tilde{\bm{u}}\mid \bm{u}) &= I \left[ G(\Tilde{\bm{u}}) \leq b_l \right]\left[\prod_{i=1}^{n_d} p_{\text{local},i}(\tilde{u}_i\mid u_i)\right]\\
        \label{eqn:global_complete_tpdf_intrepid}
        p_{\text{global}}(\tilde{\bm{u}}\mid {\bm{u}}) &= I \left[ G(\Tilde{\bm{u}}) \leq b_l \right]\left[\alpha(\Tilde{\bm{u}},\bm{u})q_\text{global} (\Tilde{\bm{u}} \mid \bm{u})+\left[1-\int{\alpha(\bm{v},\bm{u})q_{\text{global}}(\bm{v}\mid \bm{u})}\,d\bm{v} \right]\delta(\Tilde{\bm{u}}-\bm{u})\right]
\end{align}


The authors of ~\cite{chakroborty2026intrepid} show that the exploratory transitions introduced by the global proposal allow the Intrepid sampler to sample from multimodal PDFs more effectively than traditional MH samplers.



\subsection{Cyclic component-wise Intrepid MCMC}

Despite its effectiveness in sampling from multimodal distributions, the original Intrepid sampler has two features that limit its suitability for direct application to reliability estimation.

The first limitation is that the MH steps associated with $p_{\text{global}}(\cdot\mid \cdot)$ rely on $n_d$-dimensional proposal PDFs. As $n_d$ increases, the acceptance rate decreases substantially, leading to an insufficient number of accepted global exploratory moves~\cite{chakroborty2026intrepid}. This, in turn, severely limits the sampler’s ability to explore the state space. To alleviate this issue, we propose a component-wise version of Intrepid MCMC in the standard normal space, analogous to the modified Metropolis algorithm proposed by Au and Beck~\cite{au2001estimation}. 

Consider a random vector $\bm{U}=(U_{1},U_{2},\cdots,U_{n_d})$ whose Cartesian components $U_{i}$ are independent standard normal random variables. Let us denote the hyperspherical components as $\bm{U}^\text{hyp}=(R, \Theta_{1},\Theta_{2},\cdots,\Theta_{n_d-1})$. Following the derivation in~\cite{chakroborty2026intrepid}, we have,

\begin{equation}
\label{eqn:hyperspherical_pdf}
    f_{\bm{U}^\text{hyp}}(\bm{u}^\text{hyp})=(2\pi)^{-n_d/2}r^{n_d-1}\exp(-r^2/2)\prod_{i=1}^{n_d-2}\sin^{n_d-i-1}(\theta_{i}),
\end{equation}
where the hyperspherical components are independent having marginal PDFs given by
\begin{equation}
\label{eq:hyperspherical_components}
\begin{aligned}
f_R(r) 
&\propto r^{n_d-1}\exp\left(-\frac{r^2}{2}\right), 
&& r \ge 0, \\[6pt]
f_{\Theta_{i}}(\theta_i) 
&\propto \sin^{\,n_d-i-1}(\theta_i), 
&& \theta_i \in [0,\pi], \quad i = 1,\dots,n_d-2, \\[6pt]
f_{\Theta_{i}}(\theta_{i}) 
&= \frac{1}{2\pi}, 
&& \theta_{i} \in [0,2\pi], \quad i = n_d-1.
\end{aligned}
\end{equation}
This enables the implementation of component-wise MH transitions on each of the hyperspherical components. 
To do so, we define the following proposal PDFs for each of the components
\begin{equation}
\label{eq:hyperspherical_proposals}
\begin{aligned}
q_R(\Tilde r \mid r) 
&= \mathcal {N}(r,\sigma_R), \text{where } \sigma_R=\sqrt{n_d-\left(\frac{\sqrt2\Gamma\left(\frac{n_d+1}{2}\right)}{\Gamma\left(\frac{n_d}{2}\right)}\right)^2} \\[6pt]
q_{\Theta_{i}}(\Tilde{\theta}_{i}\mid \theta_{i})
&=\text{Uniform}(0,\pi), \text{for } i=1,2,\cdots,n_d-2 \\[6pt]
q_{\Theta_{i}}(\Tilde{\theta}_{i}\mid \theta_{i})
&=\text{Uniform}(0,2\pi), \text{for } i=n_d-1.
\end{aligned}
\end{equation}

These component-wise proposals are applied to the hyperspherical components with the target PDFs indicated in Eq.~\eqref{eq:hyperspherical_components}. Therefore, when transformed back to Cartesian components, the sample is distributed according to $\phi(\bm{u})$ as in the original subset simulation algorithm.

The component-wise and complete transition PDFs constructed by the proposals in Eq.~\eqref{eq:hyperspherical_proposals} are as follows.

\begin{align}
    \label{eqn:global_radial_tpdf}
    p_{\text{global},r}(\tilde{r}\mid r) &= \alpha(\Tilde{r},r)q_{R}(\tilde r\mid r) + \left[1-\int{\alpha(v,r)q_{R}(v\mid r)}\,dv \right]\delta(\tilde r-r)\\
    \label{eqn:global_angular_tpdf}
    p_{\text{global},i}(\tilde{\theta}_i\mid \theta_i) &= \alpha(\Tilde{\theta}_i,\theta_i)q_{\Theta_{i}}(\tilde \theta_{i}\mid \theta_{i}) + \left[1-\int{\alpha(v_i,\theta_i)q_{\Theta_{i}}(v_{i}\mid \theta_{i})}\,dv_i \right]\delta(\tilde \theta_{i}-\theta_{i})\\
    \label{eqn:global_complete_tpdf}
    p_{\text{global}}(\tilde{\bm{u}}\mid \bm{u}) &= I \left[ G(\Tilde{\bm{u}}) \leq b_l \right]\left[p_{\text{global},r}(\tilde{r}\mid r)\prod_{i=1}^{n_d-1} p_{\text{global},i}(\tilde{\theta}_i\mid \theta_i)\right]
\end{align}

As with Au and Beck's modified Metropolis transition $p_{\text{local}}(\cdot\mid \cdot)$ (Eq.~\eqref{eqn:local_complete_tpdf}), which updates the Cartesian components using proposals in Eq.~\eqref{eq:local_proposal} targeting the one-dimensional marginals of the parent distribution $\phi(\bm{u})$, $p_{\text{global}}(\cdot\mid \cdot)$ also requires only one performance function computation per MCMC step. In this case, the hyperspherical components are similarly updated using the proposals in (Eq.~\eqref{eq:hyperspherical_proposals}) targeting the marginals of $f_{\bm{U}^{\text{hyp}}}\left(\bm{u}^\text{hyp}\right)$, and the indicator function is evaluated only after constructing the complete candidate sample, as indicated in Eq.~\eqref{eqn:global_complete_tpdf}.

The second limitation is that the Intrepid Markov transition PDF is defined as the mixture of the comprising transition PDFs $p_{\text{local}}(\cdot\mid \cdot)$ and $p_{\text{global}}(\cdot\mid \cdot)$, as defined in Eq.~\eqref{eq:original_intrepid_mtd}. Algorithmically, this amounts to transitioning the current state $\bm{u}^{(k)}$ to the subsequent sample $\bm{u}^{(k+1)}$ using the global transition PDF with a certain probability $\beta$ or the local transition PDF with probability $1-\beta$. 
However, executing global moves only with a prescribed probability still limits the extent of exploration, resulting in suboptimal performance for reliability estimation. 

To promote more thorough exploration, instead of a mixture, we combine $p_{\text{local}}(\cdot\mid \cdot)$ and $p_{\text{global}}(\cdot\mid \cdot)$ in a cyclic fashion. Algorithmically, this entails first transitioning the current state $\bm{u}^{(k)}$ to an intermediary state $\bm{\check{u}}^{(k)}$ through the global transition PDF $p_{\text{global}}\left(\bm{\check{u}}^{(k)}\mid \bm{u}^{(k)}\right)$ and then transitioning $\bm{\check{u}}^{(k)}$ to the desired sample $\bm{u}^{(k+1)}$ using the local transition PDF $p_{\text{local}}\left(\bm{u}^{(k+1)}\mid \bm{\check{u}}^{(k)}\right)$. The PDF for the overall transition from $\bm{u}^{(k)}$ to $\bm{u}^{(k+1)}$
is then obtained as the following composition
\begin{equation}
    \label{eq:Intrepid_mtd}
    p_{\text{CI}}\left(\bm{u}^{(k+1)}\mid \bm{u}^{(k)}\right)=\int{p_{\text{local}}\left(\bm{u}^{(k+1)}\mid \check{\bm{u}}^{(k)}\right)p_\text{global}\left(\check{\bm{u}}^{(k)}\mid\bm{u}^{(k)}\right)}\, d\check{\bm{u}}^{(k)}
\end{equation}

Due to the cyclic composition of the constituent transition PDFs, generating one MCMC sample requires two performance function evaluations, one associated with executing $p_{\text{global}}\left(\check{\bm{u}}^{(k)} \mid \bm{u}^{(k)}\right)$ and the next associated with running $p_{\text{local}}\left(\bm{u}^{(k+1)} \mid \check{\bm{u}}^{(k)}\right)$. However, numerical results of the illustrative examples in Section \ref{section: numerical_illustrations} indicate that the additional computational cost is justified by the enhanced exploratory capability. Proof that the cyclic intrepid transition PDF in Eq.~\eqref{eq:Intrepid_mtd} possesses the correct stationary distribution is shown in Appendix~\ref{subsection:proof}.

Importantly, both the local and global transition PDFs as implemented in the present study employ component-wise MH steps. They may, in principle, employ any valid MCMC algorithm that samples from the target conditional PDF. The algorithm 
to implement the cyclic component-wise Intrepid sampler designed to sample from a conditional standard normal PDF $\phi(\bm{u}\mid{F_l})$ is shown in Algorithm~\ref{alg:mcintrepid_revised}.

\newlist{steps}{enumerate}{1}
\setlist[steps]{label=\textbf{Step \arabic*.}, leftmargin=*, itemsep=1ex}

\refstepcounter{algocounter}
\begin{tcolorbox}[
title=Algorithm \thealgocounter: Cyclic component-wise Intrepid MCMC,
colback=white,
colframe=black,
boxrule=0.5pt,
breakable]
\label{alg:mcintrepid_revised}

\textbf{Input:}
Performance function $G(\cdot)$, threshold $b_l$, initial state $\bm u_0$, and required number of samples $N_{s}$.\\

For $k=0,1,\ldots,N_{s}-2$:

\begin{steps}[leftmargin=1.75cm]
    \item \textbf{Global hyperspherical move.}
    
    \begin{enumerate}[label=\alph*),leftmargin=0pt]
        \item Convert $\bm u^{(k)}$ to hyperspherical coordinates
        $\left(r^{(k)},\theta_{1}^{(k)},\ldots,\theta_{n_d-1}^{(k)}\right)$.
        \item Draw a proposal sample $\Tilde{r} \sim q_R\left(\Tilde{r}\mid r^{(k)}\right)$ and compute the MH acceptance probability $\alpha_R(r^{(k)},\Tilde r)=\min\left[1,\frac{f_R(\Tilde{r})q_R\left(r^{(k)}\mid\Tilde{r}\right)}{f_R\left(r^{(k)}\right)q_R\left(\Tilde{r}\mid r^{(k)}\right)}\right]$. 
        \item Generate $v\sim \text{Uniform}(0,1)$. If $v\leq \alpha_r$, set $\bar{r}=\Tilde{r}$, else, set $\bar{r}=r^{(k)}$.
        \item For $i=1,2,\cdots,n_d-1$, draw proposal samples $\Tilde{\theta}_{i} \sim q_{\Theta_{i}}\left(\Tilde{\theta}_{i}\mid \theta_{i}^{(k)}\right)$ and compute the MH acceptance probabilities, $\alpha_{\Theta_{i}}\left(\theta_{i}^{(k)},\Tilde{\theta}_{i}\right)=\min\left[1,\frac{f_{\Theta_{i}}(\Tilde{\theta}_{i})q_{\Theta_{i}}\left(\theta_{i}^{(k)}\mid \Tilde{\theta}_{i}\right)}{f_{\Theta_{i}}\left(\theta_{i}^{(k)}\right)q_{\Theta_{i}}\left(\Tilde{\theta}_{i}\mid \theta_{i}^{(k)}\right)}\right]$. 
        \item Generate IID samples $w_j\sim\text{Uniform}(0,1),j=1,2,\cdots,n_d-1$. For each $j$, if $w_j\leq\alpha_j$, set $\bar{\theta}_{j}^{(k)}=\Tilde{\theta}_{j}$, else, set $\bar{\theta}_{j}^{(k)}=\theta_{j}^{(k)}$.
        \item Define $\bm{\bar{u}}^{\text{hyp},(k)}= \left(\bar r^{(k)},\bar\theta_{1}^{(k)},\cdots,\bar\theta_{n_d-1}^{(k)}\right)$ and its equivalent Cartesian representation $\bm{\bar{u}}^{(k)}= \left(\bar u_{1}^{(k)},\bar u_{2}^{(k)},\cdots,\bar u_{n_d-1}^{(k)}\right)$.
        \item If $G(\bm{\bar{u}}^{(k)})\leq b_l$, set $\check{\bm{u}}^{(k)}=\bm{\bar{u}}^{(k)}$. Else, set $\check{\bm{u}}^{(k)}=\bm{u}^{(k)}$. 
    \end{enumerate}



    \item \textbf{Local Cartesian move.}

    \begin{enumerate}[label=\alph*),leftmargin=0pt]
        \item For $i=1,2,\cdots,n_d$, draw proposal samples $\Tilde{u}_{i} \sim q_{U_{i}}\left(\Tilde{u}_{i}\mid \check{u}_{i}^{(k)}\right)$ and compute the MH acceptance probabilities, $\alpha_{U_{i}}\left(\check{u}_{i}^{(k)},\Tilde{u}_{i}\right)=\min\left[1,\frac{f_{U_{i}}(\Tilde{u}_{i})q_{U_{i}}\left(\check{u}_{i}^{(k)}\mid \Tilde{u}_{i}\right)}{f_{U_{i}}\left(\check{u}_{i}^{(k)}\right)q_{U_{i}}\left(\Tilde{u}_{i}\mid \check{u}_{i}^{(k)}\right)}\right]$.
        \item Generate IID samples $z_j\sim\text{Uniform}(0,1),j=1,2,\cdots,n_d$. For each $j$, if $z_j\leq\alpha_j$, set $\check{u}_{j}^{(k+1)}=\Tilde{u}_{j}$, else, set $\check{u}_{j}^{(k+1)}=\check{u}_{j}^{(k)}$.
        \item Define $\bm{\check{u}}^{(k+1)}=\left(\check u_1^{(k+1)},\check u_2^{(k+1)},\cdots,\check u_{n_d}^{(k+1)}\right)$.
        \item If $G(\bm{\check{u}}^{(k+1)})\leq b_l$, set $\bm{u}^{(k+1)}=\bm{\check{u}}^{(k+1)}$. Else, set $\bm{u}^{(k+1)}=\bm{\check{u}}^{(k)}$. 
    \end{enumerate}


\end{steps}

\textbf{Output:}
$\left\{\bm u^{(k)}\right\}_{k=0}^{N_{s}-1}$.

\end{tcolorbox}

\section{Subset Simulation with Intrepid Exploration (SuSIE)}
\label{section: proposed_method}

This section discusses the procedure to compute the probability of failure using subset simulation with the proposed cyclic component-wise Intrepid MCMC sampler and its statistical properties. 

\subsection{SuSIE Algorithm}
The proposed algorithm differs from the original subset simulation~\cite{au2001estimation} in two ways.
The first difference is that cyclic component-wise Intrepid MCMC sampler is used instead of the modified MH algorithm to draw samples from the conditional distribution $\phi(\bm{u}\mid F_l)$. The second difference is that, at each conditional level, a single Markov chain of length $n_s$ is initiated from a seed chosen uniformly at random from the set of $n_sp_0$ samples. This is in contrast to the original subset simulation method where $n_sp_0$ separate chains are generated from each seed, with each chain of a much shorter length $=1/p_0$. The disadvantage of running several short Markov chains is that if the chain states currently reside in a relatively unimportant region, then, in order to draw representative samples from the conditional PDF, all $n_sp_0$ chains must execute successful transitions to the important mode. We observed from numerical experiments that this is generally not achieved in practice. In contrast, if a single chain is initiated and it currently resides in an unimportant region, only one successful transition to the dominant mode is required to correctly draw samples from the conditional PDF. 
In the absence of geometric features that mislead subset simulation, the seeds at each intermediate level are expected to populate the relevant failure regions in proportion to their importance. In such cases, 
running one chain from each seed propagates samples that are already approximately representative of the conditional PDF.

Algorithm~\ref{alg:susie} describes the procedure to estimate the failure probability using subset simulation with Intrepid exploration (SuSIE).

\refstepcounter{algocounter}
\begin{tcolorbox}[title=Algorithm \thealgocounter: Subset Simulation with Intrepid Exploration (SuSIE), colback=white, colframe=black, boxrule=0.5pt,breakable]
\label{alg:susie}
\textbf{Given:}
Performance function $G(\cdot)$, intermediate failure probability $p_0$, number of samples per level, $n_s$, such that $n_sp_0$ is an integer.

\begin{steps}
    \item Set intermediate failure level counter $k=1$ and generate $n_s$ IID samples $\bm{u}^{(1,k)},\bm{u}^{(2,k)},\cdots,\bm{u}^{(n_s,k)} \sim \phi(\bm{u})$.
    \item Compute $G(\bm{u}^{(i,k)})$ for $i=1,2,\cdots,n_s$ and arrange the samples in ascending order of their performance function values. Let the resulting sequence be $\bar{\bm{u}}^{(1,k)},\bar{\bm{u}}^{(2,k)},\cdots,\bar{\bm{u}}^{(n_s,k)}$.
    \item Define the intermediate threshold $b_k=G(\bm{\bar{u}}^{(n_sp_0,k)})$ and hence the intermediate failure region $F_k=\{\bm{u}\in\mathbb{R}^{n_d}:G(\bm{u})\leq b_k\}$. If $b_k \leq 0$, go to step 6.
    \item Choose a sample uniformly at random from the set $\left\{\bar{\bm{u}}^{(n_sp_0,k)},\bar{\bm{u}}^{(n_sp_0+1,k)},\cdots,\bar{\bm{u}}^{(n_s,k)}\right\}$, say $\bar{\bm{u}}$. With $\bar{\bm{u}}$ as the seed, generate $n_s$ samples from $\phi({\bm{u}}\mid F_k)$ using the component-wise Intrepid MCMC sampler (Algorithm 1). Let the generated MCMC samples be $\bm{u}^{(1,k)},\bm{u}^{(2,k)},\cdots,\bm{u}^{(n_s,k)}$.
    \item Set $k \rightarrow k+1$ and go to step 2.
    \item Output the probability of failure estimate $\hat{P}_F=(p_0)^{k-1}\left(\frac{1}{n_s}\sum_{i=1}^{n_s} I\left[G(\bm{u}^{(i,k)})\leq0\right]\right)$
\end{steps}
\end{tcolorbox}

\subsection{Statistical properties of the SuSIE estimator}
\label{section: statistical_properties}

Following the arguments presented by Au and Beck~\cite{au2001estimation} for characterizing the bias and coefficient of variation (CoV) of the original subset simulation estimator, the following can be shown for the proposed estimator:
\begin{enumerate}
    \item The fractional bias of the proposed estimator is bounded above as follows.
    \begin{equation}\label{eqn:bias_sus}\left|\mathbb{E}\left(\frac{\hat{P}_F-P_F}{P_F}\right) \right|\leq\sum_{i>j}\delta_i\delta_j+\mathcal{O}(1/N)
    \end{equation}
    \item The CoV of the proposed estimator ($\delta$) can be written as follows.
    \begin{equation}
    \label{eq:cov_decomposition}
        \delta=\sqrt{\sum_{l=1}^{n_f}{\delta_l^2}}
    \end{equation}
\end{enumerate}
As in the original subset simulation method~\cite{au2001estimation}, $\delta_l$ denotes the CoV of the $l^\text{th}$ conditional probability estimator assuming that the intermediate failure thresholds are fixed. In this section, we outline an overlapping batch means procedure for estimating $\delta_l$ for the proposed SuSIE estimator.

Consider the conditional probability for level $l$, which can be given by:
\begin{equation}
\label{eqn:conditional_probability_estimator}
    \hat{P}_l=\frac{1}{n_s}\sum_{i=1}^{n_s}I\left[G\left(\bm{u}^{(i,l)}\right) \leq b_l\right] \text{ for } l=1,2,\cdots,n_f
\end{equation}
where $b_l$ is the $l^\text{th}$ intermediate threshold and $\bm{u}^{(1,l)},\bm{u}^{(2,l)},\cdots,\bm{u}^{(n_s,l)}$ are samples of a single Markov chain (e.g.\ as generated using the cyclic component-wise Intrepid MCMC). If the Markov chain central limit theorem holds, then the sampling variance of $\hat{P}_l$ can be expressed as 
\begin{equation}
    \label{eq:sampling_var_clt}
    \sigma_l^2=\frac{1}{n_s}\left(\gamma_0+2\sum_{j=1}^{n_s} \frac{n_s-j}{n_s} \gamma_j\right)
\end{equation}
A discussion on the applicability of the Markov chain central limit theorem for the proposed sampler is given in Appendix~\ref{subsection:proof}.
Here, $\gamma_j$ denotes the $j$-lag autocovariance of 
the Markov chain formed by evaluating $I\left[G\left(\bm{u}^{(s,l)}\right)\leq b_l \right]$ on the generated samples $\bm{u}^l_s$. This is given as
\begin{equation}
    \label{eqn:j_lag_autocovariance}\gamma_j=\mathbb{E}\left[\left(I\left[G\left(\bm{u}^{(s,l)}\right)\leq b_l \right]-P_l\right) \left(I\left[G\left(\bm{u}^{(s+j,l)}\right)\leq b_l \right]-P_l\right)\right]
\end{equation}
and can be estimated as 
\begin{equation}
\label{eqn:j_lag_covariance_estimate}
    \hat\gamma_j=\frac{1}{n}\sum_{i=1}^{n-j}\left\{\left(I\left[G\left(\bm{u}^{(i,l)}\right)\leq b_l \right]-\hat P_l\right) \left(I\left[G\left(\bm{u}^{(i+j,l)}\right)\leq b_l \right]-\hat P_l\right)\right\}
\end{equation}

Estimating the sampling variance $\sigma_l^2$ by substituting $\hat\gamma_j$ from Eq.~\eqref{eqn:j_lag_covariance_estimate} into Eq.~\eqref{eq:sampling_var_clt} does not result in a consistent estimator if the samples are taken from a single Markov chain (or fixed number of chains)~\cite{geyer1992practical,gupta2025estimating}. The original subset simulation algorithm yields a consistent estimator because the chain length is fixed and increasing sample sizes are accommodated by sampling from additional (nominally independent) chains. Convergence thus follows the classical Central Limit Theorem. By contrast, methods such as the proposed approach that rely on a single change (or a fixed number of chains) must rely on alternate estimators that follow the Markov Chain Central Limit Theorem to achieve consistency.
Strategies for constructing consistent estimators for the sampling variance $\sigma_l^2$ for samples from a single Markov chain broadly fall into two categories: (a) Batch means methods and (b) Spectral window methods~\cite{Flegal2010}. 

In the present study, we employ the overlapping batch means (OBM) method~\cite{Flegal2010} for estimating sampling variance. In OBM, the Markov chain samples $\bm{u}^{(1,l)},\bm{u}^{(2,l)},\cdots,\bm{u}^{(n_s,l)}$ are grouped into $n_s-s+1$ overlapping batches, where $s$ is an algorithmic parameter specifying the batch size where it is recommended to set $s=\sqrt{n_s}$~\cite{Flegal2010}. The OBM estimate of sampling variance is given by
\begin{equation}
\hat\sigma_l^2=\frac{s}{(n_s-s)(n_s-s+1)}\sum_{i=1}^{n_s-s+1}\left(m_i-\hat{P}_l\right)^2,
\label{eqn:batch-wise_variance}
\end{equation}
where 
$m_i$ is the $i^{\text{th}}$ batch-wise conditional mean estimate which is defined as 
\begin{equation}
    m_i=\frac{1}{s}\sum_{j=i}^{s+i-1}I\left[G\left(\bm{u}^{(j,l)}\right)\leq b_l \right].
    \label{eqn:batch-wise_mean}
\end{equation} 
The CoV $\delta_l$ can then be estimated as $\hat\delta_l=\hat\sigma_l / \hat{P}_l$. The steps for estimating the sampling variance $\sigma_l^2$ and hence the CoV $\delta_l$ using OBM method are outlined in Algorithm~\ref{alg:overlapping_batch_means}.

\refstepcounter{algocounter}
\begin{tcolorbox}[title=Algorithm \thealgocounter: Overlapping batch means method for estimating $\delta_l$, colback=white, colframe=black, boxrule=0.5pt,breakable]
\label{alg:overlapping_batch_means}
\textbf{Given:}
\begin{itemize}
  \item Samples of the Markov chain generated from the $l^\text{th}$ intermediate failure region $\bm{u}^{(1,l)},\bm{u}^{(2,l)},\cdots,\bm{u}^{(n_s,l)}$ and their performance function values $G\left(\bm{u}^{(1,l)}\right),G\left(\bm{u}^{(2,l)}\right),\cdots,G\left(\bm{u}^{(n_s,l)}\right)$.
  \item Intermediate failure probability estimate from Eq.~\eqref{eqn:conditional_probability_estimator}. Note that $\hat{P}_l=p_0$ for $l=1,2,\cdots,n_f-1$.
  \item Batch size $s=\sqrt{n_s}$.
\end{itemize}
\vspace{1mm}
\begin{steps}
    \item Define batches $B_1=\left\{\bm{u}^{(1,l)},\bm{u}^{(2,l)},\cdots,\bm{u}^{(s,l)}\right\},B_2=\left\{\bm{u}^{(2,l)},\bm{u}^{(3,l)},\cdots,\bm{u}^{(s+1,l)}\right\},\cdots,B_{n_s-s+1}=\left\{\bm{u}^{(n_s-s+1,l)},\bm{u}^{(n_s-s+2,l)},\cdots,\bm{u}^{(n_s,l)}\right\}$.
    \item For each batch $B_i$, $i=1,2,\cdots,n_s-s+1$, compute and store the batch-wise mean from Eq.~\eqref{eqn:batch-wise_mean}.
    \item Compute the OBM estimate of the sampling variance from Eq.~\eqref{eqn:batch-wise_variance}
    \item Output the estimate of the CoV $\hat\delta_l=\frac{\hat\sigma_l}{\hat{P}_l}$.
\end{steps}
\end{tcolorbox}

Finally, after computing $\hat\delta_l$ for each conditional probability estimate $\hat P_l$, $l=1,2,\cdots,n_f$, the CoV for the failure probability estimator $\hat P_F$ can be estimated using Eq.~\eqref{eq:cov_decomposition}. 

\section{Numerical illustrations}
\label{section: numerical_illustrations}
To demonstrate the improvement in the estimation quality of SuSIE compared to existing state-of-the-art subset simulation algorithms, we consider four illustrative examples that each have multiple failure regions or rapidly changing performance functions. 
In each example, the performance function is first transformed into the standard normal space, and the failure probability is subsequently estimated using subset simulation, with $n_s=500,1000,2000,5000, \text{and } 10000$ samples per level, intermediate failure probability $p_0=0.1$, and equipped with the following MCMC samplers:
\begin{enumerate}
    \item[S1:] Modified Metropolis algorithm, originally proposed by Au \& Beck for subset simulation~\cite{au2001estimation}.
    \item[S2:] Conditional sampling in U-space, proposed by Papaioannou \textit{et al.,}~\cite{papaioannou2015mcmc}.
    \item[S3:] Replica exchange sampler, proposed by Sharma and Manohar~\cite{sharma2023modified} to address the difficulties characterized by features such as multiple failure regions and rapid changes in performance function values.
    \item[S4:] Proposed cyclic component-wise Intrepid sampler (SuSIE).
\end{enumerate}

For sampler S1, S3, and the local transition PDF in S4 the component-wise proposal PDF is taken to be the one-dimensional Gaussian PDF with unit standard deviation centered at the current state of the Markov Chain. 
The radial and angular proposal PDFs for sampler S4 follow Algorithm 1. The correlation parameter for sampler S2 is set to 0.8, as per~\cite{papaioannou2015mcmc}. Also, S3 requires an initial exploratory run which is allocated approximately 10\% of the total computational budget. Beyond exploring the input space, the preliminary run is also used to fix the intermediate failure regions, which is used in the main run to estimate reliability. This differs from samplers S1, S2, and S4, which use the entire computational budget to directly estimate the failure probability. 
Finally, the results from each method are compared with those from direct Monte Carlo simulations using $10^8$ samples. 

For each illustrative example, the following results are reported for each sampler and each value of $n_s$:
\begin{enumerate}
    \item $\bar{P}_F$: Average value of the failure probability estimates across 1000 independent trials.
    \item $\bar{\delta}$: Empirical CoV of the failure probability estimator from 1000 independent trials. For the proposed sampler S4, we also include in parentheses the CoV estimated from a single run of the algorithm using the overlapping batch means (OBM) method.
    \item $\bar{N}_{\text{calls}}$: Average number of performance function evaluations required in a single subset simulation run from 1000 independent trials rounded to the nearest integer. 
    \item $\bar{\Delta}=\bar{\delta}\sqrt{\bar{N}_{\text{calls}}}$: The unit CoV, which quantifies the variance reduction achieved by each method~\cite{au2007application,schueller2007benchmark}. A lower value of unit CoV indicates a larger variance reduction.
\end{enumerate}

Importantly, all examples considered here depict one or more of the geometric challenges outlined in Section \ref{subsection:challenges}, i.e.: (a) several, possibly disconnected, regions of failure, all of which may not contribute equally to failure probability, or (b) discontinuities or sharp changes in performance function values. The results, therefore, depict the ability (or lack thereof) of various MCMC samplers to contend with these geometric features. We also note that three of the four examples considered here involve low-dimensional standard normal spaces (ranging from 2 to 7 dimensions). Even in low dimensions, the results show that samplers S1 and S2 are generally unable to detect or sample the relevant failure regions, while samplers S3 and S4 are able to sample from the regions of interest in these lower-dimensional cases. The final example comprises reliability estimation over a 1003-dimensional standard normal space. Here, in addition to samplers S1 and S2, even sampler S3 fails to detect the failure region of interest. In this case, only the proposed sampler S4 can identify and adequately sample from the relevant failure region.

\subsection{Example 1: Piece-wise linear function}
\label{subsection:ex1}
Consider the following performance function defined in the two-dimensional standard normal space~\cite{breitung2019geometry}:
\begin{equation}
\label{eqn:breitung_performance_function}
\begin{aligned}
    G(\bm{U}) &= \min{\left( G_1(\bm{U}), G_2(\bm{U})\right)}, \text{where,}\\
    G_1(\bm{U}) &=
    \begin{cases}
    4-U_1, & U_1 > 3.5 \\
    0.85-0.1U_1,  & U_1 \leq 3.5
    \end{cases}\\
    G_2(\bm{U}) &=
    \begin{cases}
    0.5-0.1U_2, & U_2 > 2 \\
    2.3-U_2,  & U_2 \leq 2
    \end{cases}
\end{aligned}
\end{equation}

\begin{figure}[!ht]
\centering
\includegraphics[scale=0.075]{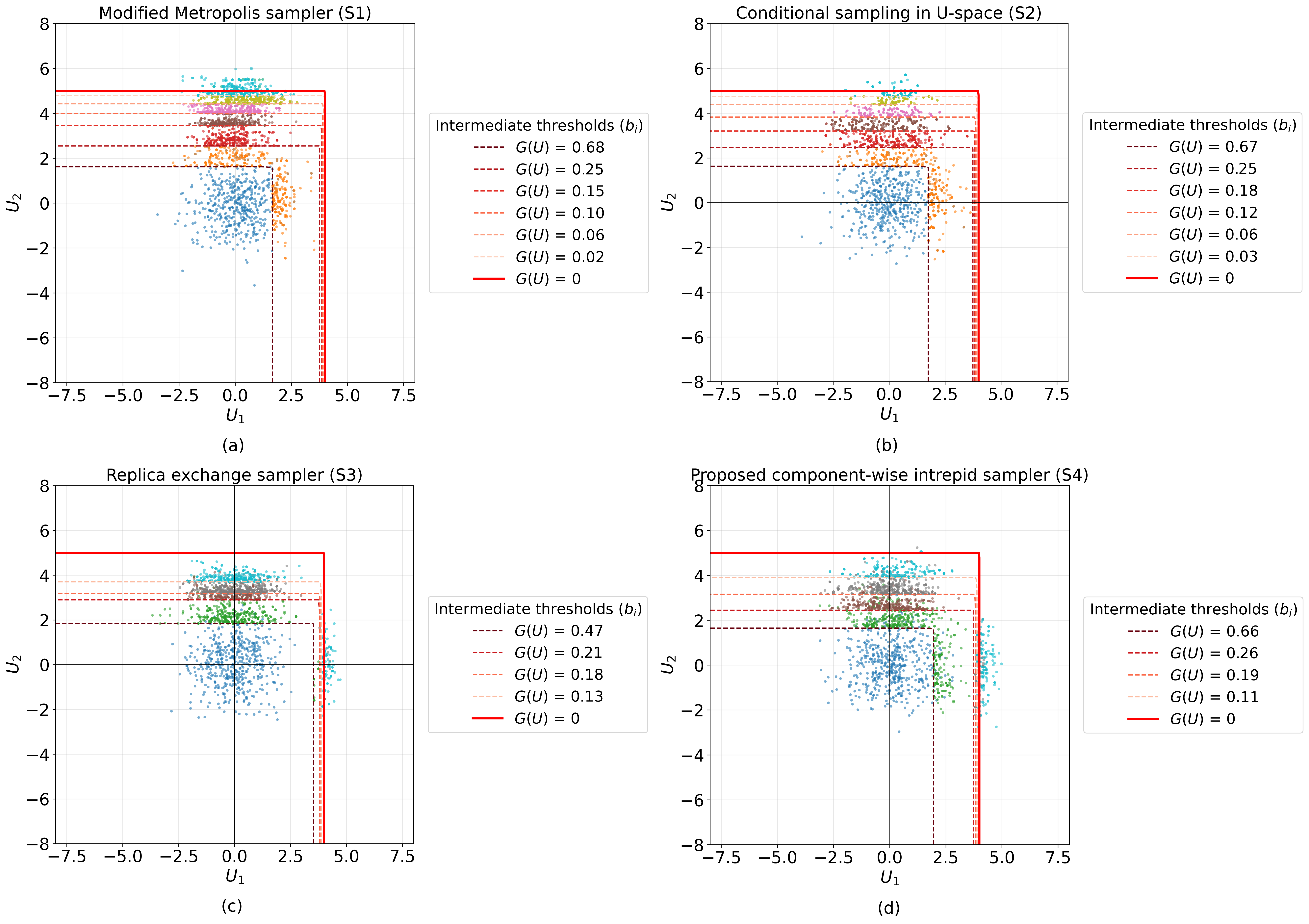}  
\caption{Example 1: Propagation of conditional samples towards the failure region for $n_s=500$ generated by: (a) Modified Metropolis sampler (S1), (b) conditional sampling in U-space (S2), (c) replica exchange sampler (S3), and (d) proposed component-wise Intrepid sampler (S4).}
\label{fig:sampleseg1}
\end{figure}

In this example, the dominant failure region corresponds to large values of $U_1$, but this region is difficult to discover due to a change in the direction of sharp decline in the performance function, as can be observed in the contours of the performance function shown in Figure \ref{fig:sampleseg1}. Consequently, the dominant mode of the conditional PDFs in the subset simulation procedure shifts to a different region at later intermediate levels.
The underlying sampler, therefore, must detect this change in mode and adapt the sampling accordingly. If it fails to identify the shift in the dominant mode, it will be misled into continued sampling from the original, less important mode, which results in substantially underestimating the failure probability. 

Figures~\ref{fig:sampleseg1}a,b show that samplers S1 and S2 are misled into sampling from the less important failure region. 
Since these samplers do not possess a mechanism of exploring the state space, they are unable to detect the shift in the dominant mode as the intermediate thresholds decrease. Meanwhile, samplers S3 and S4 are capable of recognizing the transition in the important mode and shift sampling to the important mode at later conditional levels. 

\begin{figure}[!ht]
\centering
\includegraphics[scale=0.67]{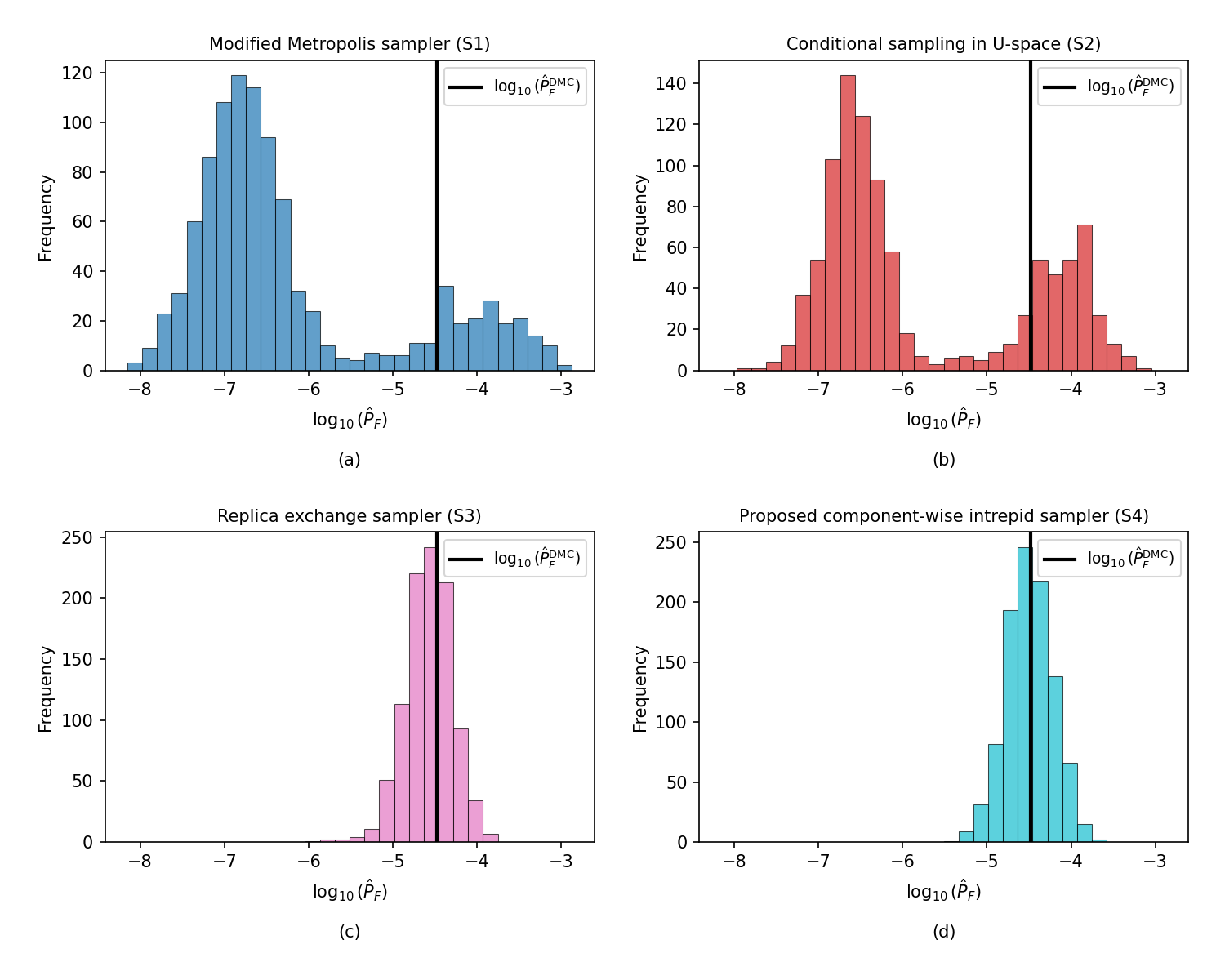}  
\caption{Example 1: Histogram of log of the failure probability estimates $\log_{10}(\hat{P}_F)$ for 1000 trials with $n_s=500$ produced by: (a) Modified Metropolis sampler (S1), (b) conditional sampling in U-space (S2), (c) replica exchange sampler (S3), and (d) proposed component-wise Intrepid sampler (S4). The vertical black line indicates $\log_{10}(\hat{P}^{MC}_F)$.}
\label{fig:histeg1}
\end{figure}

Accordingly, Figures \ref{fig:histeg1}a,b show that the histograms of failure probability estimates from 1000 trials with samplers S1 and S2 are bimodal. The dominant mode, which underestimates $P_F$ by 2 to 3 orders of magnitude corresponds to the algorithm missing the important failure mode. The second, smaller mode, corresponds to cases where the sampler finds the important mode and therefore estimates $P_F$ more accurately. Meanwhile, Figures \ref{fig:histeg1}c,d show that samplers S3 and S4 consistently identify the correct failure mode and therefore produce estimates that are consistently closer to the true failure as identified using MC simulation.

The numerical results presented in Table \ref{tab:eg1} show that, although the average values of the failure probability estimates from samplers S1 and S2 ($\bar{P}_F$) are reasonably close to the direct Monte Carlo estimate ($\hat{P}^{\text{MC}}_F$), the empirical CoVs as well as the unit CoVs ($\bar{\Delta}$) are unacceptably high, indicating a highly unstable estimator. 
The high values of the CoVs are a direct result of the fact that, in many trials, these samplers miss the region of failure that contributes the most to the failure probability. The replica exchange sampler (S3) and the proposed sampler (S4) have comparatively lower values of CoV and unit CoV. This, likewise, directly results from the sampler's ability to more consistently detect and produce samples from the region of interest. 
Sampler S3, which requires fewer performance function calls than S4, is able to discover the relevant failure region by investing approximately 10\% of the total computational effort to initially explore the low-dimensional standard normal space. 
Meanwhile, instead of an initial exploratory phase, sampler S4 executes an exploration step, and a corresponding extra model evaluation, at every Markov transition. Despite the increased computational effort for sampler S4, the numerical results show that the CoV and unit CoV values are close to those of S3. 

Thus, in this low-dimensional example, S3 and S4 are able to generate accurate and stable estimates of the failure probability while S1 and S2 are not, with S3 moderately outperforming the proposed method S4. However, the proposed method does not require any change to the implementation of subset simulation or the treatment of the generated samples outside of replacing modified Metropolis with the component-wise Intrepid MCMC sampler. On the other hand, S3 requires a specialized initial exploratory run, which changes its implementation complexity compared to S4 and standard subset simulation.

\begin{table}[!ht]
\centering
\caption{Numerical results for Example 1. $\bar{P}_F$ and $\bar{\delta}$ are average failure probaiblity and CoV from 1000 independent trials. $\bar{N}_{\text{calls}}$ is the average number of performance function evaluations from 1000 independent trials. $\bar{\Delta}=\bar{\delta}/\bar{N}_{\text{calls}}$.}
\label{tab:eg1}
\begin{tabular}{cccccc}
\toprule
$n_s$ & Method & $\bar{P}_F$ & $\bar{N}_{\text{calls}}$ & $\bar{\delta}$ & $\bar{\Delta}$ \\
\midrule
\multirow{4}{*}{500} & S1 & $3.70\times10^{-5}$ & $2775$ & $3.21$ & $169.50$ \\
                     & S2 & $3.65\times10^{-5}$ & $2861$ & $2.11$ & $113.29$ \\
                     & S3 & $3.21\times10^{-5}$ & $3094$ & $0.65$ & $36.67$ \\
                     & S4 & $3.87\times10^{-5}$ & $4267$ & $0.67 (0.65)$ & $44.41$ \\
\midrule
\multirow{4}{*}{1000} & S1 & $3.55\times10^{-5}$ & $5193$ & $2.19$ & $158.14$\\
                      & S2 & $3.49\times10^{-5}$ & $5394$ & $1.42$ & $104.60$\\
                      & S3 & $3.16\times10^{-5}$ & $6122$ & $0.43$ & $33.85$ \\
                      & S4 & $3.58\times10^{-5}$  & $8546$ & $0.47 (0.43)$ & $44.18$ \\
\midrule
\multirow{4}{*}{2000} & S1 & $3.32\times10^{-5}$ & $9752$ & $1.59$ & $157.80$ \\
                      & S2 & $3.29\times10^{-5}$ & $10037 $ & $1.05$ & $105.24$ \\
                      & S3 & $3.21\times10^{-5}$ & $12150$ & $0.30$ & $33.22$ \\
                      & S4 & $3.45\times10^{-5}$ & $17092$ & $0.32 (0.34)$ & $43.04$\\
\midrule
\multirow{4}{*}{5000} & S1 & $3.36\times10^{-5}$ & $22508$ & $1.04$ & $157.46$ \\
                      & S2 & $3.27\times10^{-5}$ & $23621$ & $0.66$ & $101.66$\\
                      & S3 & $3.19\times10^{-5}$ & $30130$ & $0.18$ & $31.88$ \\
                      & S4 & $3.26\times10^{-5}$ & $42740$ & $0.19 (0.20)$ & $40.80$ \\
\midrule
\multirow{4}{*}{10000} & S1 & $3.29\times10^{-5}$ & $43246$ & $0.74$ & $155.62$ \\
                       & S2 & $3.15\times10^{-5}$ & $46432$ & $0.47$ & $102.08$ \\
                       & S3 & $3.18\times10^{-5}$ & $60268$ & $0.12$ & $31.89$ \\
                       & S4 & $3.25\times10^{-5}$ & $85480$ & $0.13 (0.15)$ & $39.22$ \\
    \midrule
    \multicolumn{6}{l}{Direct Monte Carlo estimate with $10^8$ samples, $\hat{P}_F^{MC} = 3.27\times10^{-5}$} \\
    \bottomrule
\end{tabular}
\end{table}

\subsection{Example 2: Vibration of a two-DOF system}
\label{subsection:ex2}
In this example, we consider a two-DOF mass-spring-dashpot system driven by sinusoidal excitation as shown in Figure \ref{fig:structeg2} (studied previously by Sharma and Manohar~\cite{sharma2023modified}). Spring stiffnesses $K_1$ and $K_2$ are assumed to be IID random variables with $K_1,K_2 \sim \text{Lognormal}\left(\mu=2.5\times10^5 \text{ N/m}, \text{CoV}=0.2\right)$. The masses $M_1=M_2=2000 \text{ kg}$ and the modal damping ratios $\eta_1=\eta_2=0.02$ are considered to be deterministic. The mass $M_2$ is subjected to excitation $F(t)=2000\sin{(11t)} \text{ N}$, yielding the performance function defined as follows
\begin{equation}
\label{eqn:eg2_performance_function}
    g(K_1,K_2) = 0.024 - \max_{t\in[0 \text{ s},20 \text{ s}]}{x_1(t)}
\end{equation}
\textbf{\begin{figure}[!ht]
\centering
\includegraphics[scale=0.5]{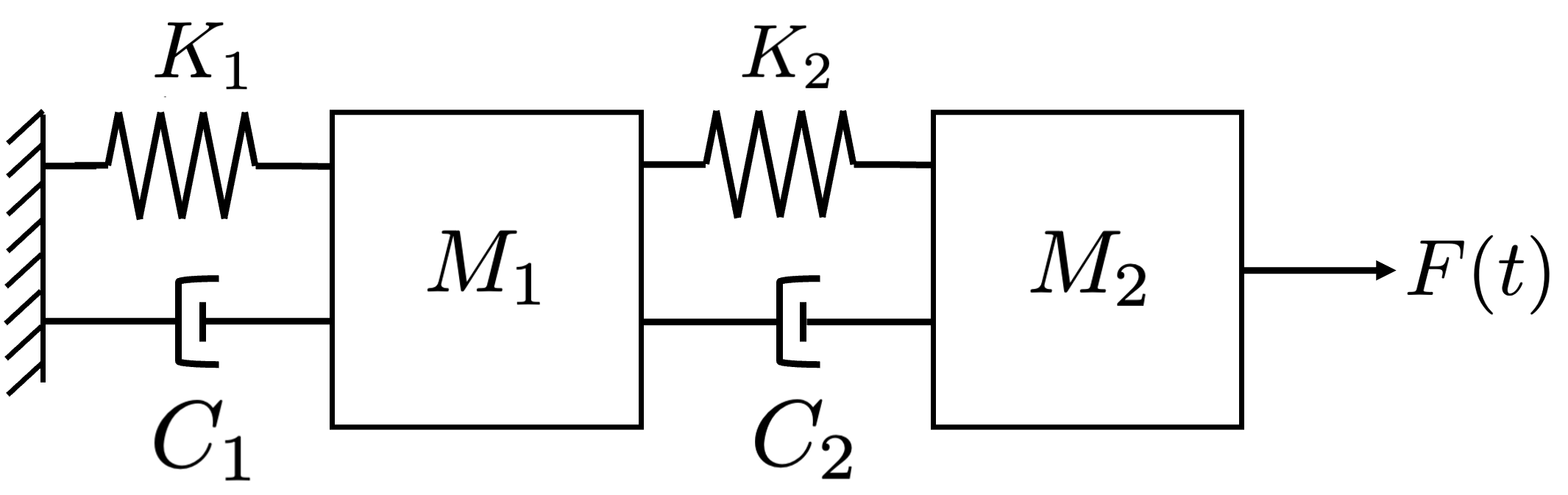}  
\caption{Two-degree-of-freedom mass-spring-damper system under sinusoidal excitation considered in Example 2.}
\label{fig:structeg2}
\end{figure}}

\begin{figure}[!ht]
\centering
\includegraphics[scale=0.075]{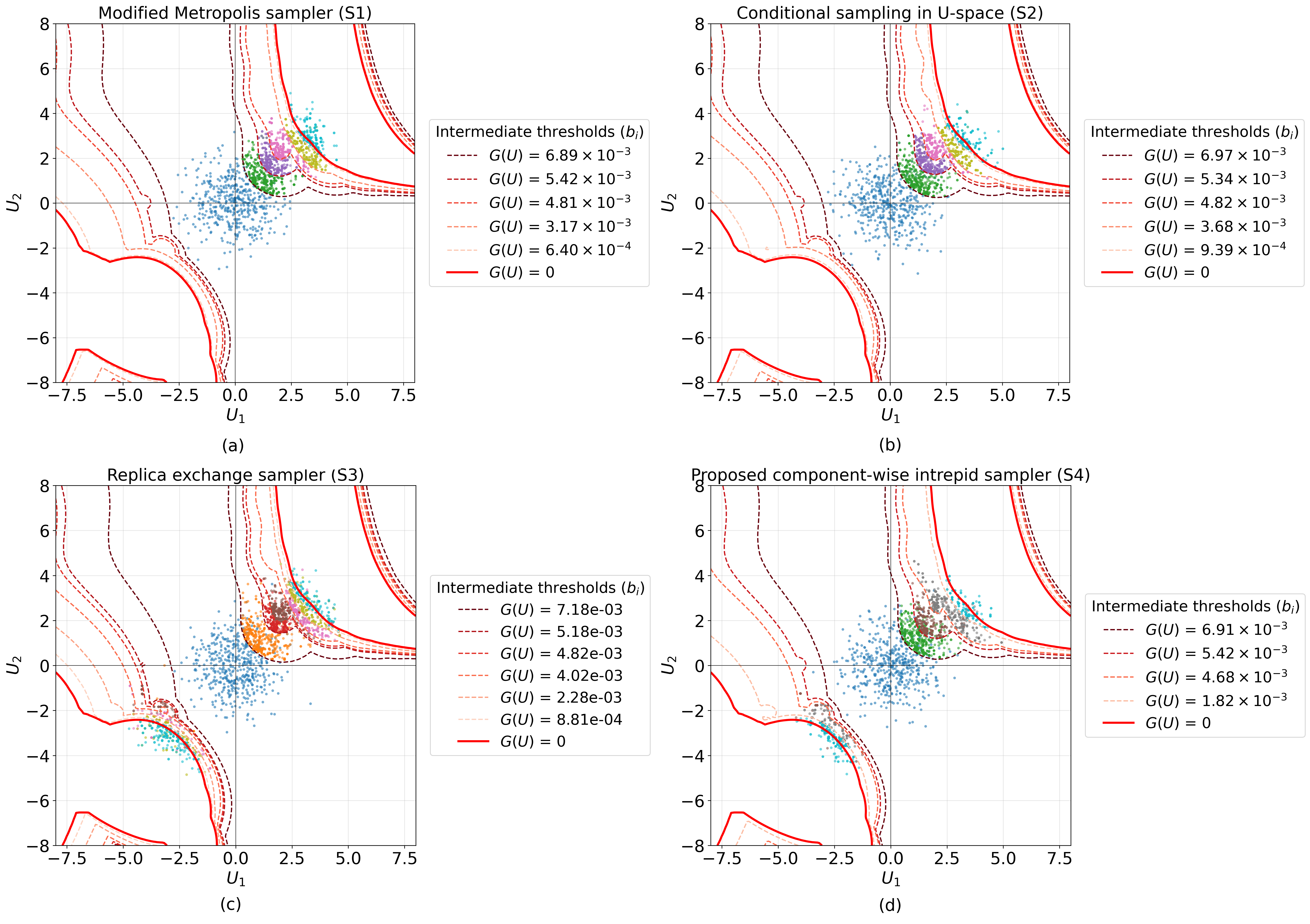}  
\caption{Example 2: Propagation of samples towards the failure region for $n_s=500$ generated by: (a) Modified Metropolis sampler (S1), (b) conditional sampling in U-space (S2), (c) replica exchange sampler (S3), and (d) proposed component-wise Intrepid sampler (S4).}
\label{fig:sampleseg2}
\end{figure}

This problem has two disconnected failure regions that contribute comparably to the failure probability. Due to the irregular geometric structure of the intermediate failure regions (as visualized in Figure~\ref{fig:sampleseg2}), MCMC samplers may be unable to propagate samples into both of the relevant failure regions. Here, it is crucial for the sampler not only to identify both modes but also to execute sufficient Markov transitions between them so that sampling is proportional to their relative contributions to the failure probability. Failure to adequately traverse both modes will lead to incorrect estimates of the failure probability.

\begin{figure}[!ht]
\centering
\includegraphics[scale=0.67]{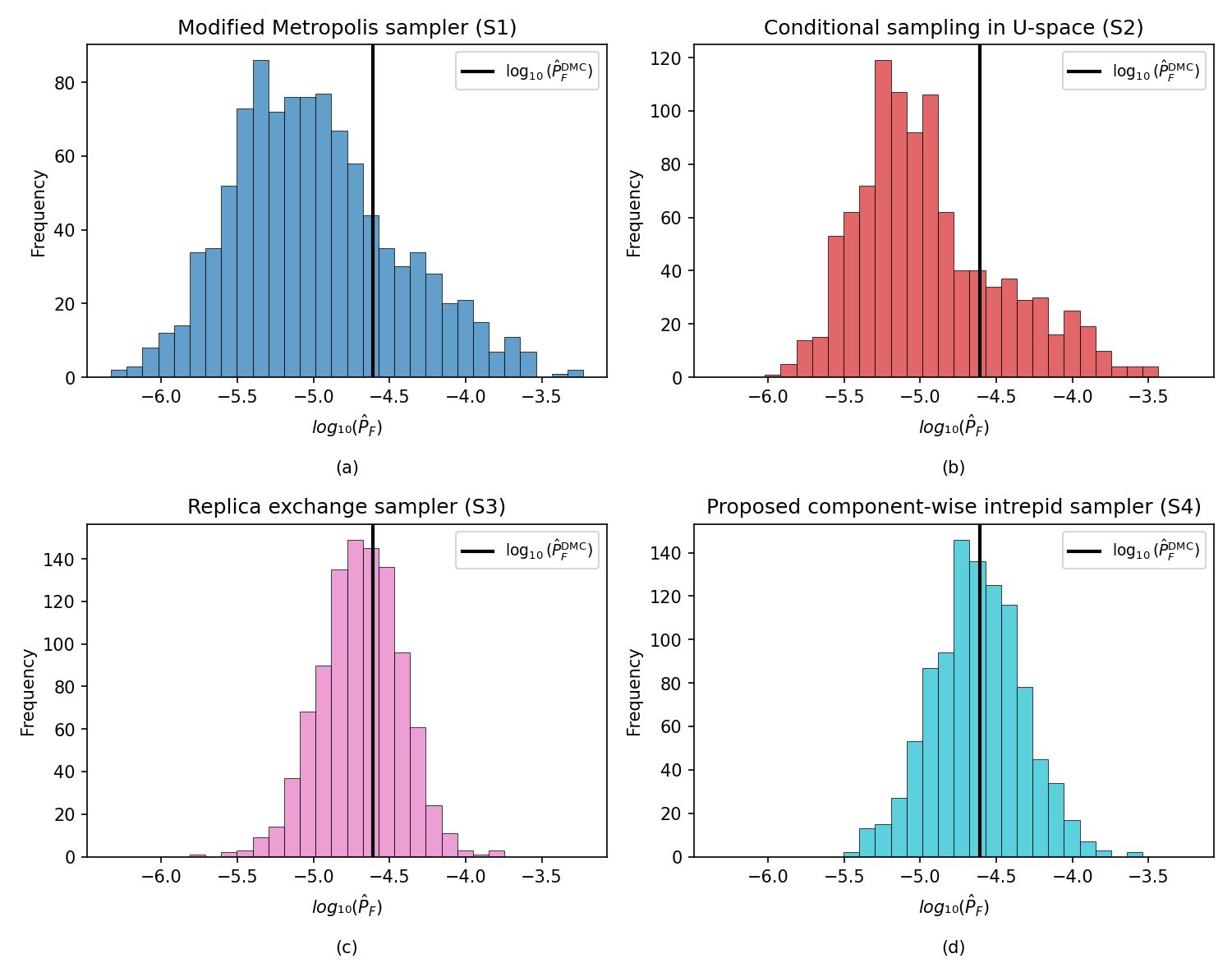}  
\caption{Example 2: Histogram of log of the failure probability estimates $\log_{10}(\hat{P}_F)$ for 1000 trials with $n_s=500$ produced by : (a) Modified Metropolis sampler (S1), (b) conditional sampling in U-space (S2), (c) replica exchange sampler (S3), and (d) proposed component-wise Intrepid sampler (S4). The vertical black line indicates $\log_{10}(\hat{P}^{MC}_F)$.}
\label{fig:histeg2}
\end{figure}

In the trials shown in Figures~\ref{fig:sampleseg2}a,b, samplers S1 and S2 are only able to sample from one of the failure regions, while S3 and S4 sample from both. Accordingly, the histograms in Figures~\ref{fig:histeg2}a,b show that a significant fraction of the trials of S1 and S2 underestimate the failure probability. This contributes to the large CoV associated with samplers S1 and S2 reported in Table~\ref{tab:eg2}, since the failure probability estimates differ substantially depending on whether a particular run sampled from both or only one of the failure regions. Samplers S3 and S4, on the other hand, demonstrate improved performance with comparatively more consistent estimates of $P_F$ (Figures~\ref{fig:histeg2}c,d) and lower CoV (Table~\ref{tab:eg2}). 

\begin{table}[!ht]

\centering
\caption{Numerical results for Example 2. $\bar{P}_F$ and $\bar{\delta}$ are average failure probability and CoV from 1000 independent trials. $\bar{N}_{\text{calls}}$ is the average number of performance function evaluations from 1000 independent trials. $\bar{\Delta}=\bar{\delta}/\bar{N}_{\text{calls}}$.}
\label{tab:eg2}
\begin{tabular}{cccccc}
\toprule
$n_s$ & Method & $\bar{P}_F$ & $\bar{N}_{\text{calls}}$ & $\bar{\delta}$ & $\bar{\Delta}$ \\
\midrule
\multirow{4}{*}{500} & S1 & $2.43\times10^{-5}$ & $2274$ & $1.91$ & $91.45$\\
                     & S2 & $2.31\times10^{-5}$ & $2512$ & $1.65$ & $83.09$ \\
                     & S3 & $2.31\times10^{-5}$ & $3126$ & $0.68$ & $38.29$ \\
                     & S4 & $3.00\times10^{-5}$ & $4334$ & $0.79 (0.60)$ & $52.26$ \\
\midrule
\multirow{4}{*}{1000} & S1 & $2.35\times10^{-5}$ & $4476$ & $1.23$ & $82.81$ \\
                      & S2 & $2.22\times10^{-5}$ & $4956$ & $1.11$ & $78.34$ \\
                      & S3 & $2.39\times10^{-5}$ & $6224$ & $0.46$ & $36.30$\\
                      & S4 & $2.64\times10^{-5}$ & $8588$ & $0.48 (0.45)$ & $44.75$ \\
\midrule
\multirow{4}{*}{2000} & S1 & $2.35\times10^{-5}$ & $8762$ & $0.85$ & $79.88$ \\
                      & S2 & $2.23\times10^{-5}$ & $9690$ & $0.75$ & $73.92$ \\
                      & S3 & $2.30\times10^{-5}$ & $12418$ & $0.32$ & $36.13$\\
                      & S4 & $2.43\times10^{-5}$ & $17049$ & $0.32 (0.33)$ & $42.59$ \\
\midrule
\multirow{4}{*}{5000} & S1 & $2.30\times10^{-5}$ & $21177$ & $0.51$ & $74.96$ \\
                      & S2 & $2.26\times10^{-5}$ & $23369$ & $0.47$ & $72.24$ \\
                      & S3 & $2.32\times10^{-5}$ & $30495$ & $0.20$ & $34.96$ \\
                      & S4 & $2.40\times10^{-5}$ & $42575$ & $0.21 (0.21)$ & $43.71$ \\
\midrule
\multirow{4}{*}{10000} & S1 & $2.37\times10^{-5}$ & $41782$ & $0.37$ & $76.08$\\
                       & S2 & $2.35\times10^{-5}$ & $46099$ & $0.33$ & $72.53$ \\
                       & S3 & $2.32\times10^{-5}$ & $60415$ & $0.13$ & $33.71$ \\
                       & S4 & $2.38\times10^{-5}$ & $85148$ & $0.14 (0.14)$ & $41.91$ \\
\midrule
\multicolumn{6}{l}{Direct Monte Carlo estimate with $10^8$ samples, $\hat{P}_F^{MC} = 2.40\times10^{-5}$} \\
\bottomrule
\end{tabular}
\end{table}

We also observe in Table~\ref{tab:eg2} that as $n_s$ increases, the CoV for samplers S1 and S2 decreases significantly, indicating that they are more likely to sample from both failure regions. However, even when both regions are detected by these samplers, they are unlikely to propose Markov transitions between the modes. This lack of mixing between modes leads to an increase in CoV compared to S3 and S4, even for larger $n_s$. In contrast, samplers S3 and S4
are able to identify and propose transitions between both important modes, leading to more accurate failure probability estimates and lower CoV values.

Similar to example 1, the initial exploration phase utilized by sampler S3 appears to be sufficient to explore the two dimensional standard normal space and identify the important failure regions. The increased number of performance function calls for sampler S4 is tied to the fact that an exploration step is executed at each Markov chain step. However, despite the increased computational effort for sampler S4, once again the unit CoV values are comparable to S3 and much lower than S1 and S2.

\subsection{Example 3: Stable symmetric buckling of rigid bars}
\label{subsection:ex3}
\begin{figure}[!ht]
\centering
\includegraphics[scale=0.5]{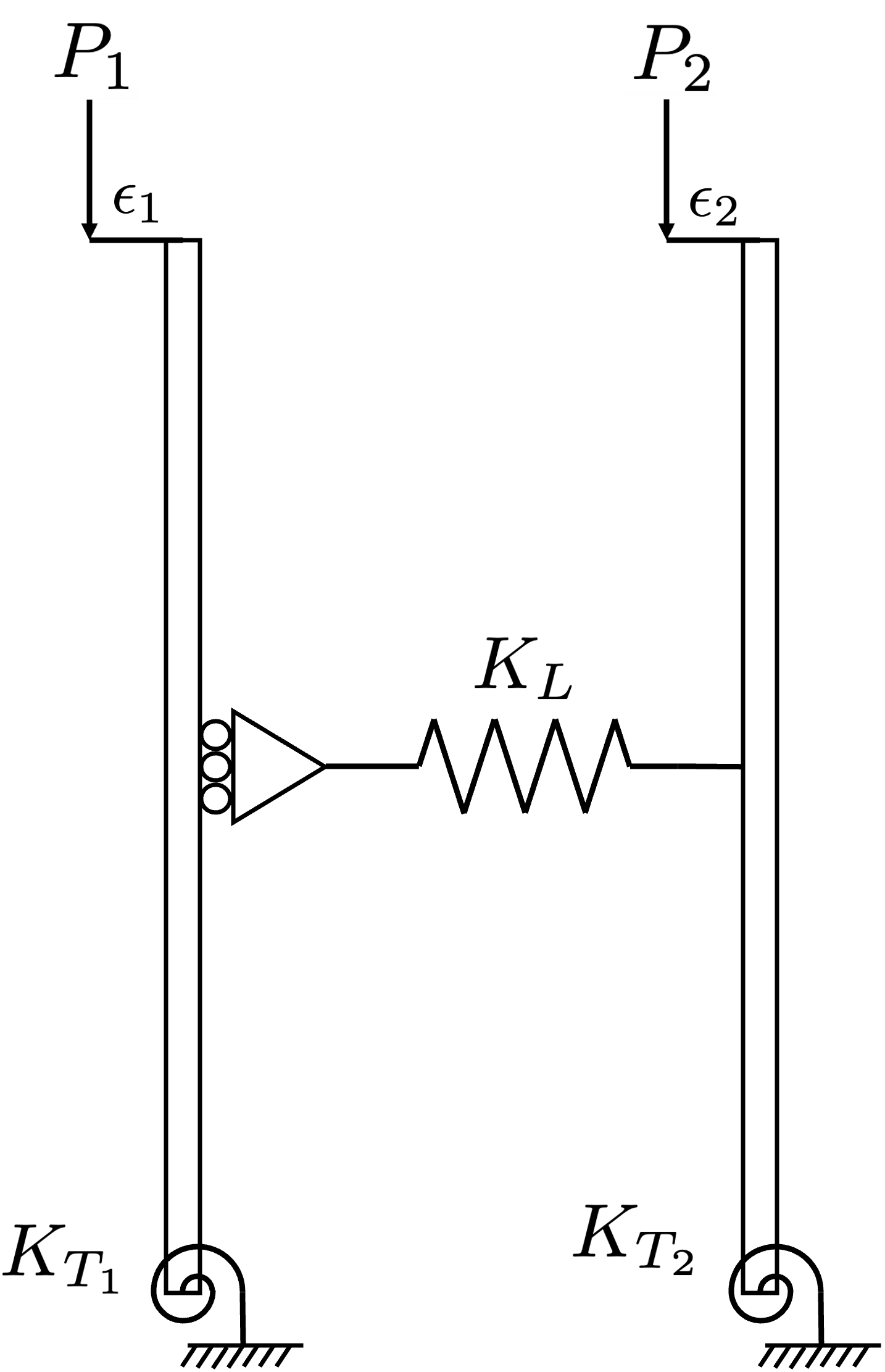}  
\caption{Example 3: Eccentrically loaded rigid bars connected by a linear spring with torsional springs at the base.}
\label{fig:structeg3}
\end{figure}

Next, we consider a structural system comprising two rigid massless bars of length $L=1$ m supported by torsional springs at the base and connected by a linear spring fixed at the midpoint of the right bar, as shown in Figure \ref{fig:structeg3} and previously considered by Chakroborty and Shields~\cite{chakroborty2025tail}. The rigid bars are subjected to eccentric downward loads, causing stable symmetric buckling. Let $\theta_1$ and $\theta_2$ denote the angular deflection of the left and right bars with respect to the vertical, respectively, taken to be positive in the counter-clockwise direction. The problem is considered to have the following seven random variables: (a) the torsional spring stiffnesses $K_{T_1},K_{T_2} \sim \text{Beta}(a=5,b=5)$ with upper and lower limits $1.7 \text{ GNm and } 2.3 \text{ GNm}$, respectively, (b) the linear spring stiffness $K_L \sim \text{Beta}(a=5,b=5)$ with upper and lower limits $0.8 \text{ GNm and } 1.2 \text{ GNm}$, (c) the eccentricities $\epsilon_1, \epsilon_2 \sim \text{Uniform}(-0.05 \text{ m},0.05 \text{ m})$, and (d) the vertical loads $P_1,P_2 \sim f_P(p)$, where $f_P(p)$ represents a shifted lognormal PDF given by 
\begin{equation}
    f_P(p)=\frac{1}{(p-1)\sigma\sqrt{2\pi}}\exp\left[-\frac{(\ln(p-1)-\mu)^2}{2\sigma^2}\right],
\end{equation} 
where $p>1$ GN, $\mu=1+\ln(0.1)$, and $\sigma=0.25$.
All random variables are assumed to be mutually independent.

\begin{figure}[!ht]
\centering
\includegraphics[width=0.78\textwidth]{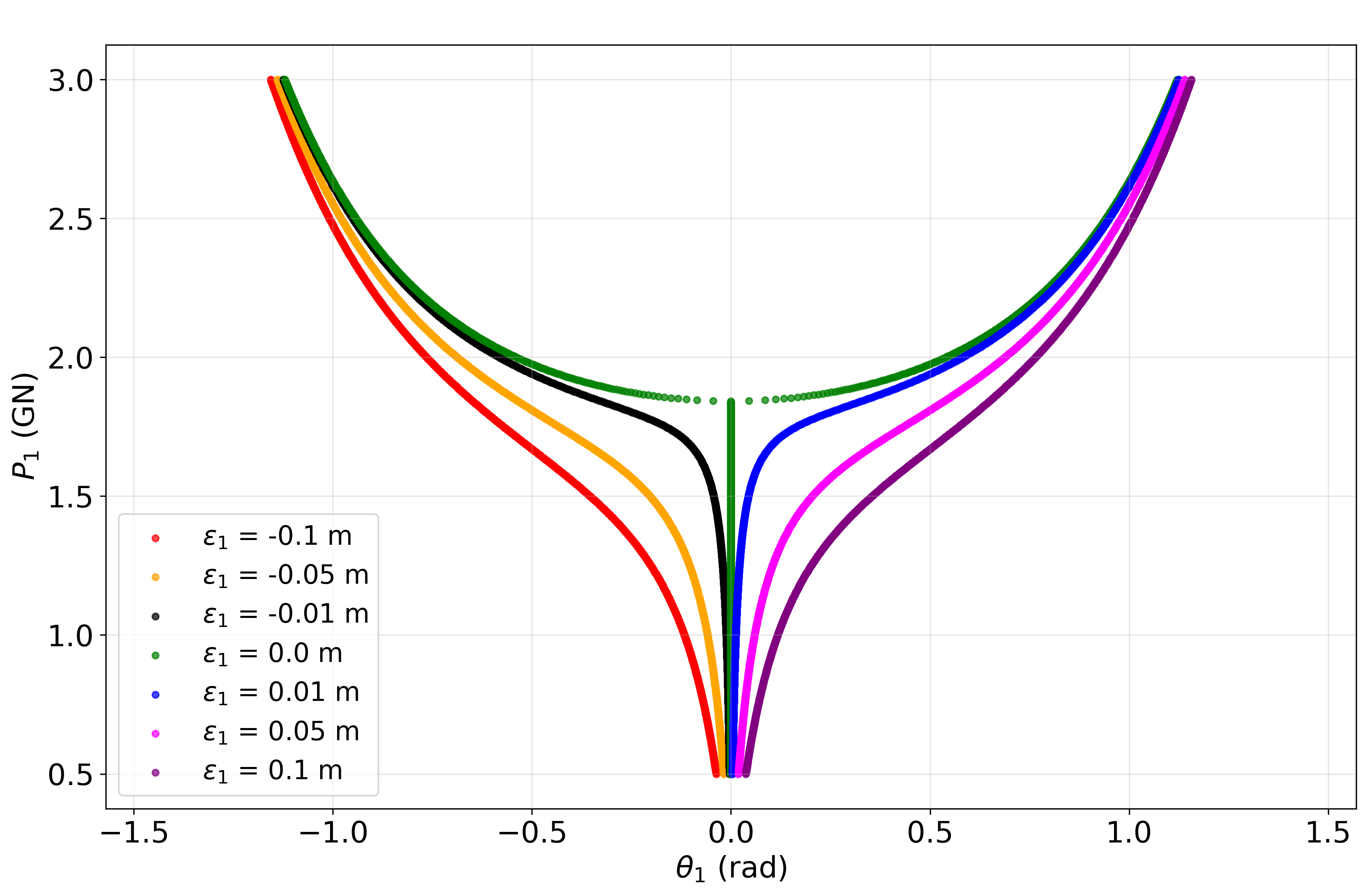}  
\caption{Example 3: Load ($P_1$) vs deflection ($\theta$) for various values of eccentricity ($\epsilon_1$) for the structure.}
\label{fig:loaddefeg3}
\end{figure}

Let $\bm X=\begin{bmatrix}
    K_{T_1} & K_{T_2} & K_L & \epsilon_1 & \epsilon_2 & P_1 & P_2 
\end{bmatrix}^\text{T}$ denote the 7-dimensional input random vector. The load-deflection diagrams of $P_1 \text{ vs } \theta_1$ for various values of $\epsilon_1$ when $P_2=1.2 \text{ GN}, K_L= 0.8 \text{ GN}, \epsilon_2= 0 \text{ m}, \text { and } K_{T_1}=K_{T_2}=1.7 \text{ GN}$ are shown in Figure \ref{fig:loaddefeg3}.
The structure is deemed to have failed if there exists a stable equilibrium state of the structure where either of bar exceeds an angular deflection of $0.2$ radians. The performance function is defined accordingly as follows:
\begin{equation}
\label{eqn:eg3_performance_function}
    g(\bm X) = 0.2 - \max{(|\theta_1|,|\theta_2|)}
\end{equation}

\begin{figure}[!ht]
\centering
\includegraphics[width=0.7\textwidth]{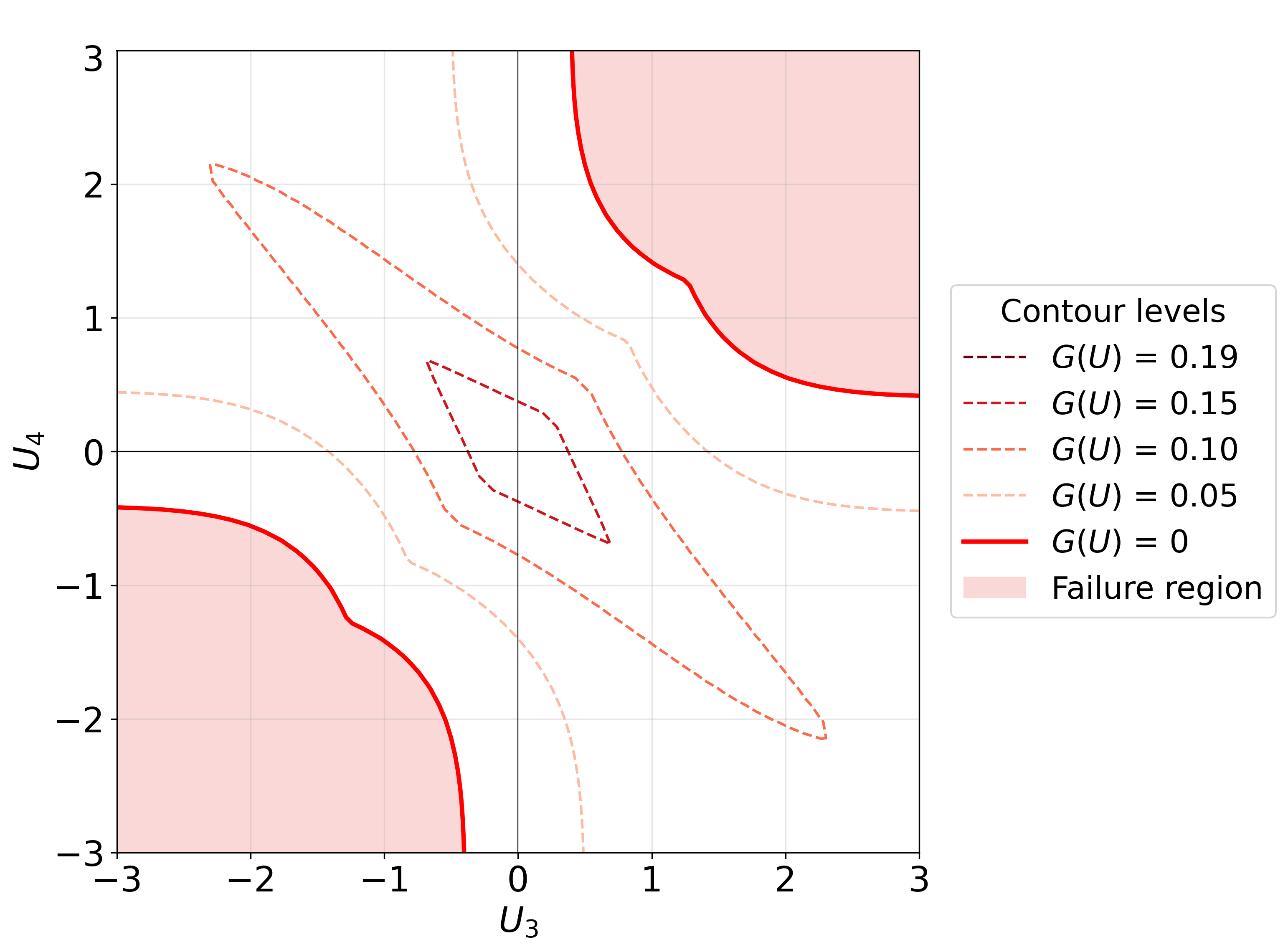}  
\caption{Example 3: 2-dimensional slice ($U_3$ vs $U_4$) of the 7-dimensional performance function $G(\bm{U})$ setting $U_1=U_2=2.5$ and $U_4=U_5=U_6=U_7=-2.5$.}
\label{fig:sliceeg3}
\end{figure}

In a simplified two-dimensional version of this problem~\cite{sharma2023modified} where only the eccentricities $\epsilon_1$ and $\epsilon_2$ were treated as uniform random variables, it was shown that the failure domain is comprised of two disconnected regions, resulting in a multimodal conditional PDF associated with each intermediate subset. Therefore, in this more complicated variant of the problem, we expect similar geometric features to lead to a multimodal conditional PDF in the 7-dimensional standard normal space. Although we cannot visualize the performance function in its entirety, we nevertheless plot the contours of a 2-dimensional slice of the 7-dimensional performance function transformed into the standard normal space, $G(\bm{U})$, in Figure~\ref{fig:sliceeg3}. Setting $U_1=U_2=2.5$ and $U_5=U_6=U_7=-2.5$, we let $U_3$ and $U_4$ be the free variables for plotting the contours, and observe disconnected failure regions in the 2-dimensional slice that suggest the existence of multimodalities in the 7-dimensional conditional PDFs.

\begin{figure}[htbp]
\centering
\includegraphics[width=\textwidth]{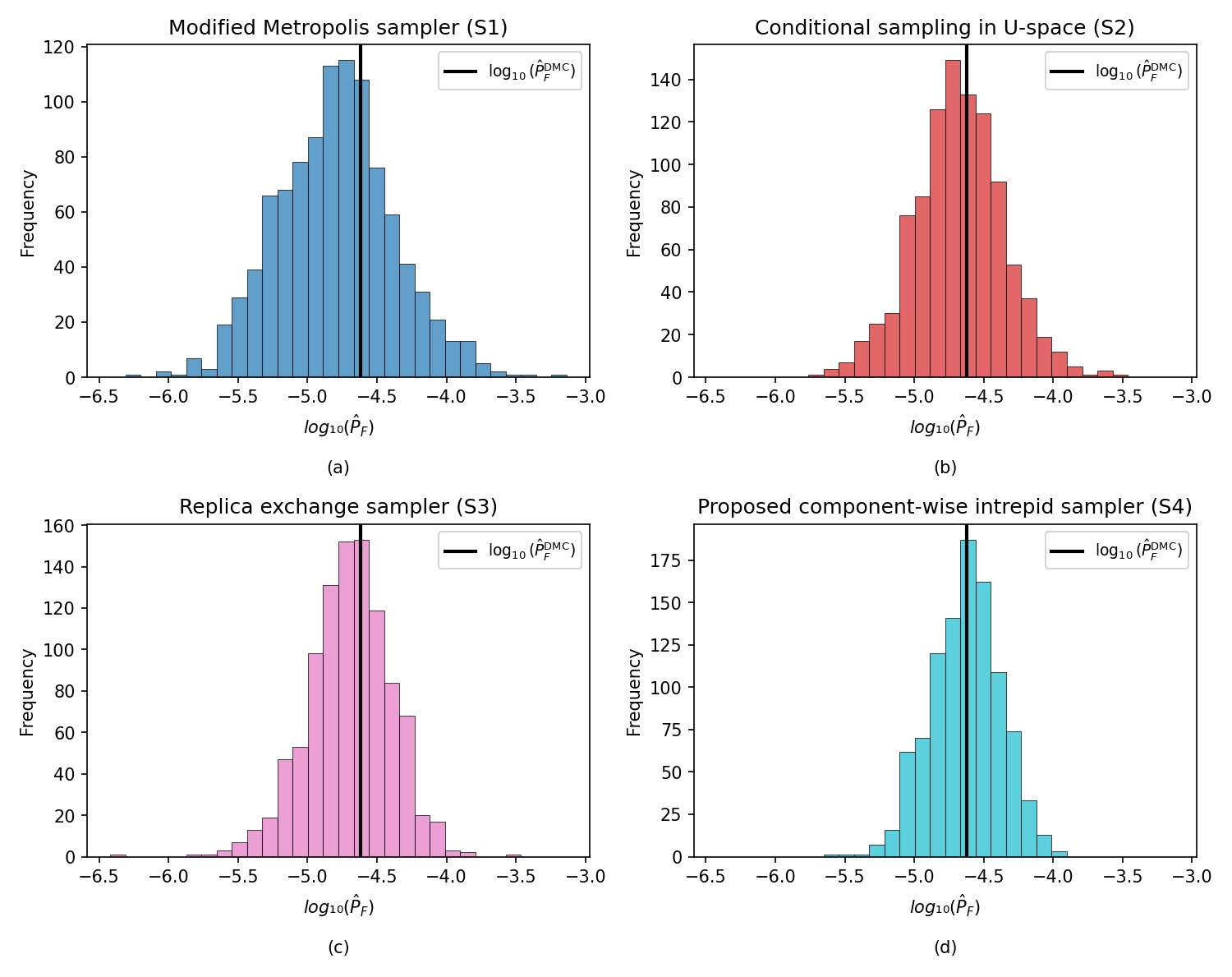}  
\caption{Example 3: Histogram of log of the failure probability estimates $\log_{10}(\hat{P}_F)$ from 1000 trials with $n_s=500$ produced by : (a) Modified Metropolis sampler (S1), (b) conditional sampling in U-space, (c) replica exchange sampler, and (d) proposed component-wise Intrepid sampler. The vertical black line indicates $\log_{10}(\hat{P}^{MC}_F)$.}
\label{fig:histeg3}
\end{figure}

Figure \ref{fig:histeg3} shows that none of the samplers in this case grossly underestimate the failure probability, implying that none of the samplers miss out on an important region. Nevertheless, numerical results in Table \ref{tab:eg3} still show that sampler S1 has larger CoV than the other methods for identical $n_s$. Again, the larger CoV arises from the fact that sampler S1 is unable to sufficiently transition samples between the modes. For sampler S2, we also note that while its CoV is comparable to sampler S4 especially for large $n_s$, it is distinctly higher for $n_s = 500$ and $n_s = 1000$. Therefore, it appears that although the mixing of sampler S2 is better than sampler S1, the lack of exploratory steps results in inferior performance compared to the proposed sampler S4 for smaller sample sizes per subset.

\begin{table}[!ht]
\centering
\caption{Numerical results for Example 3. $\bar{P}_F$ and $\bar{\delta}$ are the average failure probability and CoV from 1000 independent trials. $\bar{N}_{\text{calls}}$ is the average number of performance function evaluations from 1000 independent trials. $\bar{\Delta}=\bar{\delta}/\bar{N}_{\text{calls}}$.}
\label{tab:eg3}
\begin{tabular}{cccccc}
\toprule
$n_s$ & Method & $\bar{P}_F$ & $\bar{N}_{\text{calls}}$ & $\bar{\delta}$ & $\bar{\Delta}$ \\
\midrule
\multirow{4}{*}{500} & S1 & $2.59\times10^{-5}$ & $2422$ & $1.48$ & $73.18$\\
                     & S2 & $2.75\times10^{-5}$ & $2359$ & $0.90$ & $43.97$ \\
                     & S3 & $2.50\times10^{-5}$ & $3399$ & $0.79$ & $46.07$\\
                     & S4 & $2.73\times10^{-5}$ & $4583$ & $0.58 (0.56)$ & $39.62$ \\
\midrule
\multirow{4}{*}{1000} & S1 & $2.50\times10^{-5}$ & $4756$ & $0.94$ & $64.92$\\
                      & S2 & $2.57\times10^{-5}$ & $4635$ & $0.56$ & $38.14$ \\
                      & S3 & $2.44\times10^{-5}$ & $6788$ & $0.49$ & $41.00$ \\
                      & S4 & $2.61\times10^{-5}$ & $9047$ & $0.44 (0.45)$ & $42.14$ \\
\midrule
\multirow{4}{*}{2000} & S1 & $2.47\times10^{-5}$ & $9318$ & $0.59$ & $57.69$ \\
                      & S2 & $2.43\times10^{-5}$ & $9211$ & $0.36$ & $34.63$ \\
                      & S3 & $2.43\times10^{-5}$ & $13644$ & $0.34$ & $40.16$ \\
                      & S4 & $2.49\times10^{-5}$ & $18010$ & $0.29 (0.30)$ & $39.95$ \\
\midrule
\multirow{4}{*}{5000} & S1 & $2.41\times10^{-5}$ & $23017$ & $0.37$ & $56.27$ \\
                      & S2 & $2.41\times10^{-5}$ & $23000$ & $0.21$ & $33.21$\\
                      & S3 & $2.37\times10^{-5}$ & $33817$ & $0.21$ & $39.19$ \\
                      & S4 & $2.42\times10^{-5}$  & $44994$  & $0.18 (0.20)$ & $39.62$ \\
\midrule
\multirow{4}{*}{10000} & S1 & $2.42\times10^{-5}$ & $45990$ & $0.25$ & $55.11$ \\
                       & S2 & $2.40\times10^{-5}$ & $46000$ & $0.16$ & $34.44$ \\
                       & S3 & $2.40\times10^{-5}$ & $66853$ & $0.16$ & $41.45$ \\
                       & S4 & $2.41\times10^{-5}$  & $89988$  & $0.13 (0.15)$ & $40.31$ \\
\midrule
    \multicolumn{6}{l}{Direct Monte Carlo estimate with $10^8$ samples, $\hat{P}_F^{MC} = 2.35\times10^{-5}$} \\
\bottomrule
\end{tabular}
\end{table}

Of particular interest here is the result for sampler S3. Both the CoV and unit CoV values of S3 were on par with S4 in the previous two-dimensional examples. In this seven dimensional example, while the unit CoVs remain comparable, the sampler S3 exhibits higher CoVs across all $n_s$, with the disparity being particularly pronounced for smaller $n_s$. The original study by Sharma and Manohar~\cite{sharma2023modified} identified that the exploratory run, which enables global exploration and traversal between modes in S3, becomes insufficient in higher dimensions. This is corroborated by the results in Table~\ref{tab:eg3}. The proposed sampler S4 does not require such an exploratory run and is also able to effectively traverse between the modes in this seven-dimensional example.

\subsection{Example 4: Snap-through buckling of a von-Mises truss}
\label{subsection:ex4}
\begin{figure}[!ht]
\centering
\includegraphics[width=0.7\textwidth]{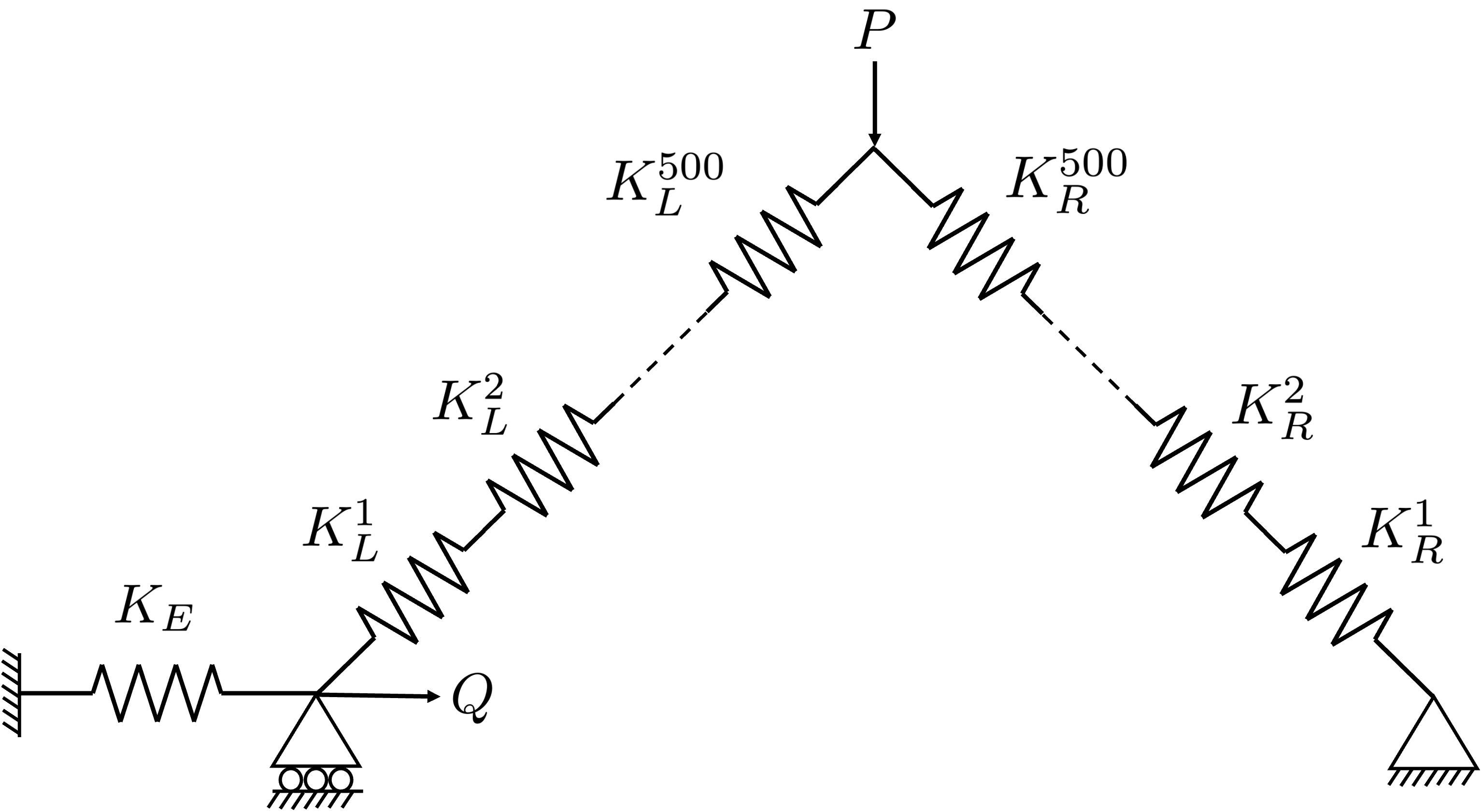}  
\caption{Example 4: A von-Mises truss prone to snap-through buckling.}
\label{fig:structeg4}
\end{figure}

Finally, consider the reliability estimation of a von-Mises truss prone to snap-through buckling as shown in Figure \ref{fig:structeg4}. Each arm of the truss is $\sqrt{2}\SI{}{m}$ long, inclined at $45^\circ$ from the horizontal, and modeled as an axially deformable bar with spatially varying axial stiffness. To model the spatial variability, we discretize each axial bar into $N_p=500$ spring elements connected in series. The stiffnesses of the spring elements are treated as random variables and denoted as $K_L^i$ and $K_R^i$ ($i=1,2,\cdots,500$), corresponding to the left and right arm, respectively. The loads $P$ and $Q$, and the stiffness of the horizontal spring attached to the left arm, $K_E$, are also treated as random variables. Table~\ref{tab:rv_params} details these random variables.

\begin{table}[!ht]
\centering
\caption{Details of random variables in Example 4. All variables follow lognormal distributions with parameters $\mu_{\ln}$ and $\sigma_{\ln}$. }
\label{tab:rv_params}
\begin{tabular}{@{}lcc@{}}
    \toprule
    Variable & Mean, CoV & $\mu_{\ln}$, $\sigma_{\ln}$ \\
    \midrule
    $P$ & $1.92$ KN, $0.05$ & $0.65$, $0.05$ \\
    $Q$ & $1.42$ KN, $0.05$ & $0.35$, $0.05$ \\
    $K_E$ & $15$ KN/m, $0.05$ & $2.70$, $0.05$ \\
    $K_L^i$, $K_R^i$, $i=1,2,\cdots,500$ & $4.99$ MN/m, $0.10$ & $8.51$, $0.10$ \\[0.75em]
    \multicolumn{3}{@{}l@{}}{Correlation coefficients: $\rho_{K_L^i,K_L^{i+1}}=0.2$ and $\rho_{K_R^i,K_R^{i+1}}=0.2$ for $i=1,2,\cdots,499$.} \\
    \bottomrule
\end{tabular}
\end{table}

\begin{figure}[!ht]
\centering
\includegraphics[width=0.5\textwidth]{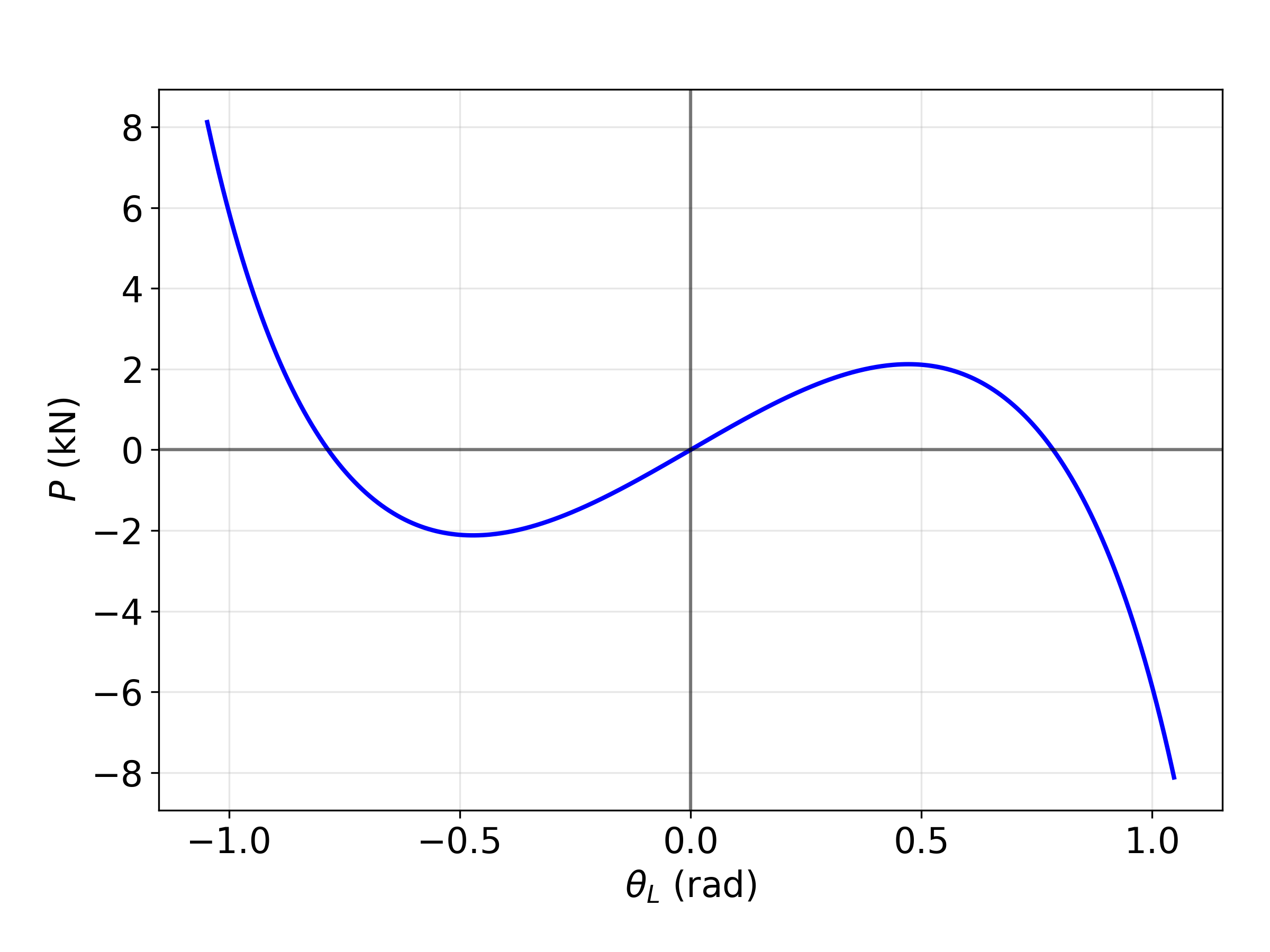}  
\caption{Example 4: Load-deflection diagram showing snap-through buckling for the von-Mises truss.}
\label{fig:loaddefeg4}
\end{figure}

Let $y$ denote the leftward deflection of the roller support, while $\theta_L$ and $\theta_R$ denote the angle that the left and right arms of the truss make with the horizontal in the deformed configuration, respectively.  Appendix~\ref{subsection:load_def_derivation} provides the derivation of the load-deflection equation for the truss. The resulting $P \text{ vs } \theta_L$ diagram is plotted in Figure~\ref{fig:loaddefeg4}, which matches the characteristic load-deflection behavior for snap-through buckling.

The original input random vector $\bm{X}=\begin{bmatrix} P & Q & K_E & K_L^1 & K_L^2 & \cdots & K_R^1 & K_R^2 & \cdots & K_R^{500} \end{bmatrix}^\text{T}$ is transformed to the standard normal space using the Nataf transformation. Failure occurs if the structure undergoes snap-through buckling or if the horizontal spring attached to the left arm undergoes a tensile deformation exceeding $y=50$ mm. This leads to the following performance function:
\begin{equation}
\label{eqn:eg4_performance_function}
    g(\bm X) = \min\left(\theta_L,0.05+y\right)
\end{equation}

It has previously been shown that traditional methods of reliability estimation may struggle with problems involving snap-through buckling~\cite{sharma2023sampling,sharma2023modified} due to a sharp decrease in the performance function value near the failure region. 
Similar to example 1, the sharp decline may cause MCMC samplers to miss the region of failure associated with snap-through buckling even in low dimensions. Thus, we expect the same difficulties to be further exacerbated in the much larger 1003-dimensional problem considered here. 

\begin{figure}[!ht]
\centering
\includegraphics[width=\textwidth]{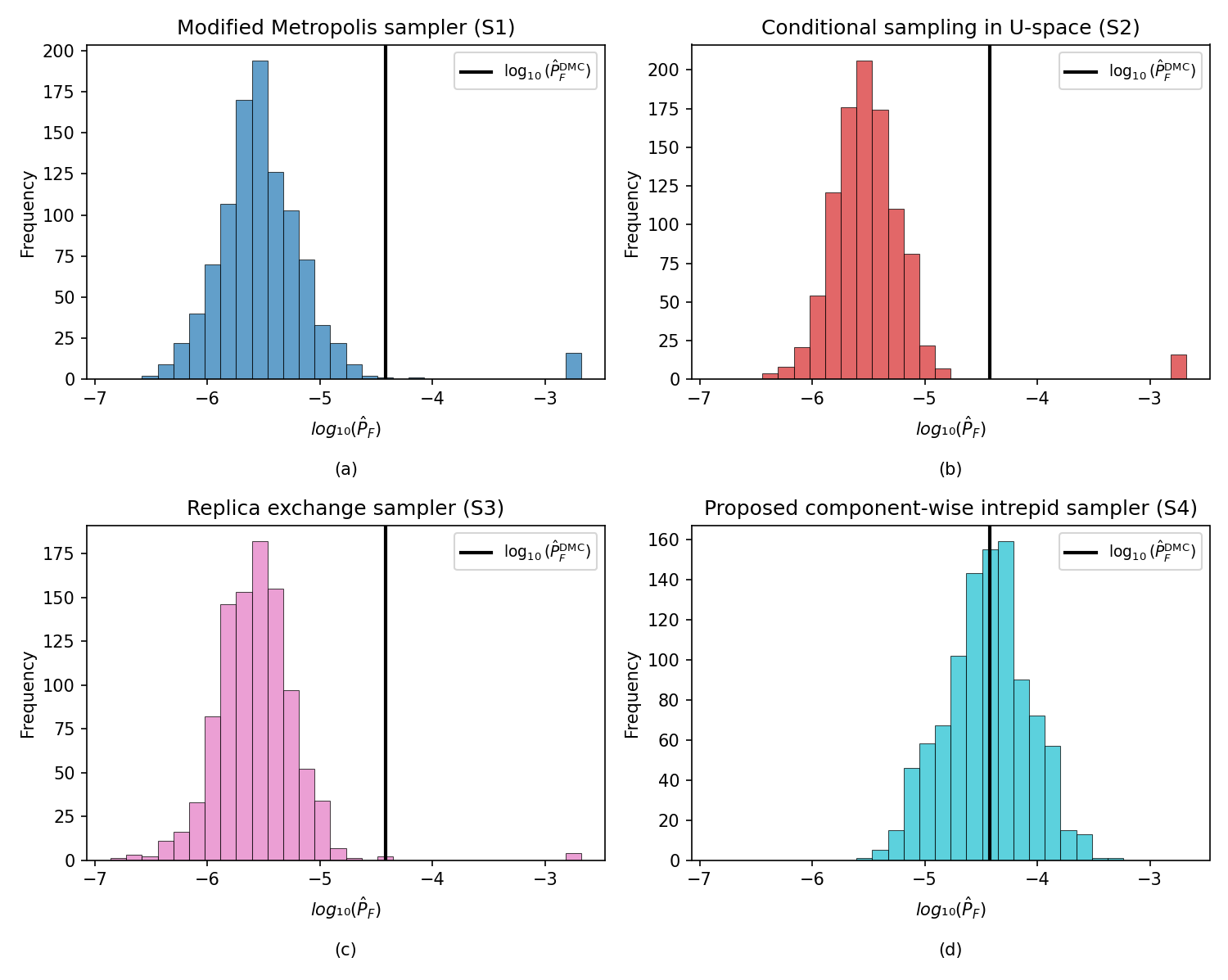}  
\caption{Example 4: Histogram of log of the failure probability estimates $\log_{10}(\hat{P}_F)$ from 1000 trials with $n_s=500$ produced by : (a) Modified Metropolis sampler (S1), (b) conditional sampling in U-space, (c) replica exchange sampler, and (d) proposed component-wise Intrepid sampler. The vertical black line indicates $\log_{10}(\hat{P}^{MC}_F)$.}
\label{fig:histeg4}
\end{figure}

Indeed, the histograms of failure probability estimates given in Figure~\ref{fig:histeg4} clearly indicate that all samplers except for S4 miss the most important failure region in the vast majority of the subset simulation trials. Accordingly, the numerical results presented in Table \ref{tab:eg4} show unacceptably high CoV and unit CoV for samplers S1, S2, and S3, while those for the proposed sampler S4 indicate stable estimates. Note that sampler S3, even with a preliminary exploration, was unable to identify and sample from the failure region of interest in this high dimensional case. Sampler S4, however, successfully identified the important region in the 1003-dimensional standard normal space, and yielded acceptable estimates of failure probability.

\begin{table}[!ht]
\centering
\caption{Numerical results for Example 4. $\bar{P}_F$ and $\bar{\delta}$ are average failure probability and CoV from 1000 independent trials. $\bar{N}_{\text{calls}}$ is the average number of performance function evaluations from 1000 independent trials. $\bar{\Delta}=\bar{\delta}/\bar{N}_{\text{calls}}$.}
\label{tab:eg4}
\begin{tabular}{cccccc}
\toprule
$n_s$ & Method & $\bar{P}_F$ & $\bar{N}_{\text{calls}}$ & $\bar{\delta}$ & $\bar{\Delta}$ \\
\midrule
\multirow{4}{*}{500} & S1 & $3.58\times10^{-5}$ & $2743$ & $6.99$ & $366.52$ \\
                    & S2 & $3.55\times10^{-5}$ & $2737$ & $7.06$ & $369.34$ \\
                     & S3 & $1.12\times10^{-5}$ & $3417$ & $11.17$ & $652.99$ \\
                     & S4 & $5.04\times10^{-5}$ & $4480$ & $0.92 (0.62)$ & $62.12$ \\
\midrule
\multirow{4}{*}{1000} & S1 & $3.36\times10^{-5}$ & $5417$ & $5.07$ & $373.26$ \\
                      & S2 & $3.33\times10^{-5}$ & $5422$ & $5.10$ & $375.88$ \\
                      & S3 & $1.63\times10^{-5}$ & $6617$ & $6.88$ & $559.87$ \\
                      & S4 & $4.78\times10^{-5}$ & $8892$ & $0.70 (0.58)$ & $66.30$ \\
\midrule
\multirow{4}{*}{2000} & S1 & $3.58\times10^{-5}$ & $10768$ & $3.43$ & $356.45$ \\
                      & S2 & $3.58\times10^{-5}$ & $10766$ & $3.43$ & $356.92$ \\
                      & S3 & $1.36\times10^{-5}$ & $13026$ & $5.12$ & $584.80$ \\
                      & S4 & $4.17\times10^{-5}$ & $17944$ & $0.45 (0.41)$ & $61.57$ \\
\midrule
\multirow{4}{*}{5000} & S1 & $3.95\times10^{-5}$ & $25961$ & $2.08$ & $336.46$ \\
                      & S2 & $3.95\times10^{-5}$ & $25961$ & $2.08$ & $336.30$\\
                      & S3 & $1.42\times10^{-5}$ & $32030$ & $3.12$ & $559.20$ \\
                      & S4 & $3.92\times10^{-5}$ & $45000$ & $0.27 (0.28)$ & $57.43$ \\
\midrule
\multirow{4}{*}{10000} & S1 & $3.88\times10^{-5}$ & $49528$ & $1.49$ & $333.61$ \\
                       & S2 & $3.88\times10^{-5}$ & $49528$ & $1.49$ & $333.72$ \\
                       & S3 & $1.32\times10^{-5}$ & $63928$ & $2.23$ & $564.60$\\
                       & S4 & $3.87\times10^{-5}$ & $90000$ & $0.18 (0.19)$ & $54.67$ \\
\midrule
    \multicolumn{6}{l}{Direct Monte Carlo estimate with $10^8$ samples, $\hat{P}_F^{MC} = 3.80\times10^{-5}$} \\
\bottomrule
\end{tabular}
\end{table}

\subsection{Comparison of coefficient of variation for similar performance function calls}
\begin{figure}[!ht]
\centering
\includegraphics[scale=0.67]{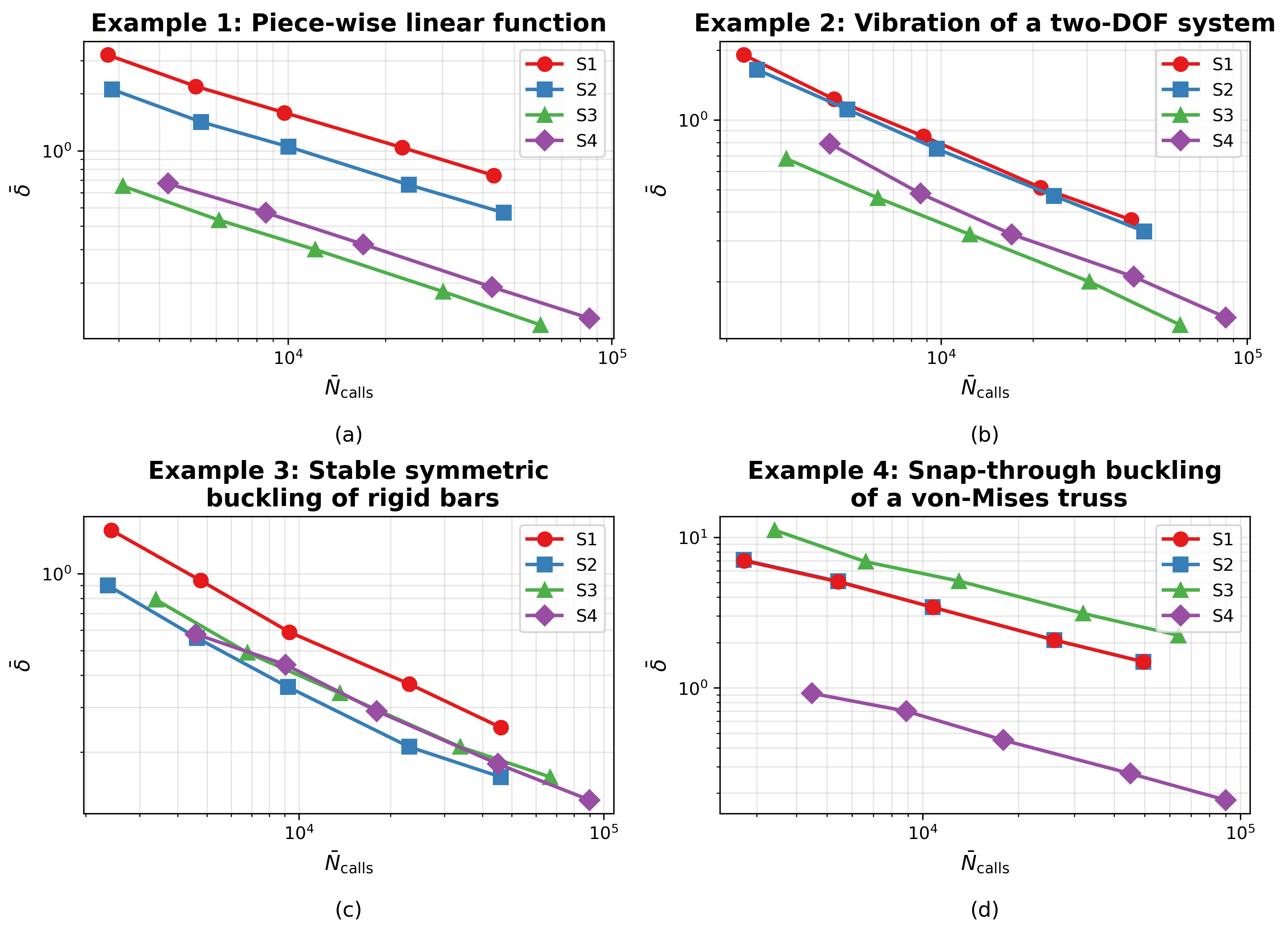}  
\caption{CoV $(\bar\delta)$ vs number of performance function calls $N_\text{calls}$ for samplers S1, S2, S3, and S4 for: (a) Example 1: Piece-wise linear function, (b) Example 2: Vibration of a two-DOF system, (c) Example 3: Stable symmetric buckling of rigid, and (d) Example 4: Snap-through buckling of a von-Mises truss.}
\label{fig:delta_vs_ncalls_loglog}
\end{figure}

In practice, reliability analyses are often performed under a fixed computational budget. Therefore, it is of interest to assess the CoV achieved by each method for a comparable number of performance function calls. Figure~\ref{fig:delta_vs_ncalls_loglog} compares the CoV for each method vs. number of function calls for the four examples considered in this section.

Consistent with the numerical results presented in the previous sections, we observe that the modified Metropolis algorithm (S1) yields relatively large CoV for a given number of performance function calls across all four examples. Although it performs better than S3 in the 1003-dimensional snap-through buckling example, the practical relevance of such a comparison is limited given that the CoV values for these methods remain unacceptably large (ranging approximately from 2 to 10). 

Sampler S2 similarly yields fairly high values of CoV for most cases. A notable exception is for the stable symmetric buckling case (Example 3), where it performs on par with samplers S3 and S4 and outperforms sampler S1. We recall that the stable symmetric buckling case involves a seven-dimensional multimodal PDF where none of the four methods missed a failure region. We also note, based on Table~\ref{tab:eg3}, that although sampler S2 has noticeably higher CoV for small $n_s$ values compared to S3 and S4, it performs comparably with them in terms of unit CoV values. This suggests that the multimodality is not severe enough to degrade the performance of S2, especially for larger values of $n_s$. In contrast, in the vibration example (Example 2) and the snap-through buckling example (Example 4), S1 and S2 performed comparably poorly, both yielding high CoV values.

Sampler S3, as noted in the previous sections, performs well in low to moderate dimensional settings (examples 1, 2, and 3) by utilizing 10\% of the total computational to explore the state space. This helps the sampler achieve acceptable values of CoV, and remain competitive with sampler S4 in these cases. However, in the high dimensional snap-through buckling case (Example 4), it produces unacceptably large CoV, and is outperformed even by samplers S1 and S2.

The proposed sampler S4 is clearly the only sampler that achieves acceptably low values of CoV in all examples. All other samplers exhibit excessively high CoV values for at least one of the examples. Notably, even with approximately 50,000 to 60,000 function calls Table~\ref{tab:eg4} shows that methods S1, S2, and S3 are unable to achieve a CoV that sampler S4 is able to attain with merely 4,480 sample calls. Additionally, sampler S4 remains competitive in all other cases (Examples 1, 2, and 3), achieving CoV values comparable to the lowest observed across S1, S2, and S3 in these examples. This demonstrates strong performance of the proposed sampler across the numerical examples considered, maintaining competitiveness in low-to moderate-dimensional settings while outperforming the other methods in the high-dimensional case.

\subsection{Limitations and future directions}
\label{section: future_directions}

The proposed sampler has been designed to contend with geometric difficulties relating to multiple failure regions or sharply varying performance functions in the standard normal space. While this remains an important class of unresolved problems, it is not the only geometric issue that can arise in reliability contexts. 
One challenge that remains is when, rather than the conditional distributions being disjoint or multi-modal, the conditional distributions are extremely narrow. 
An example of this can be observed the Black Swan problem discussed by Au and Wang~\cite{au2014engineering}, where the performance function is defined as follows:
\begin{equation}
\label{eqn:blackswan_performance_function}
        G(\bm{U}) =
        \begin{cases}
        5-U_1, & U_1 \leq 2 \\
        5-U_2,  & U_1 > 2
        \end{cases}\\
\end{equation}

\begin{figure}[!ht]
\centering
\includegraphics[scale=0.075]{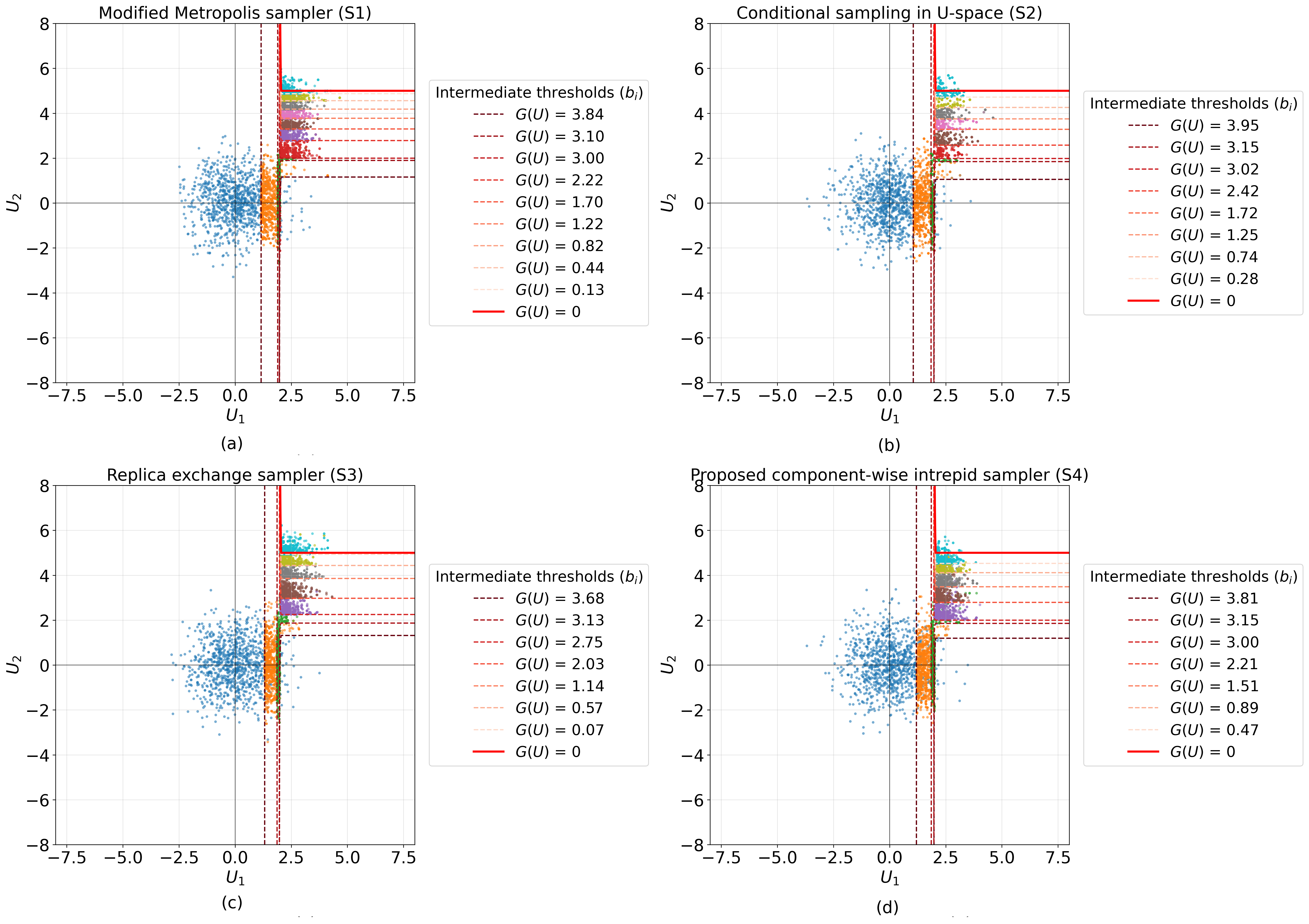}  
\caption{Black Swan problem: Propagation of samples towards the failure region for $n_s=1000$ generated by: (a) Modified Metropolis sampler (S1), (b) conditional sampling in U-space, (c) replica exchange sampler, and (d) proposed cyclic component-wise Intrepid sampler.}
\label{fig:samplesegblackswan}
\end{figure}

Figure \ref{fig:samplesegblackswan} shows the propagation of generated samples towards the failure region for all four samplers considered in Section \ref{section: numerical_illustrations}, while Table \ref{tab:egblackswan} summarizes the numerical results, for $1000$ samples per intermediate level. Samplers S1 and S2 have high CoV, while the replica exchange sampler (S3) greatly overestimates the failure probability. The proposed sampler S4 also slightly overestimates the failure probability. Its empirical CoV, although lower than that of samplers S1 and S2, is still quite high. However, despite not satisfactorily resolving this particular geometric challenge, S4 performs better than the other samplers considered here. It provides a substantially more accurate estimate than S3 and a lower empirical CoV than S1 and S2. The reason sampler S4 (and sampler S3) does not perform well in this case is because the geometric challenge here is associated with drawing samples from a conditional PDF that is extremely narrow, and not with either a rapid change in the performance function or the presence of multiple failure regions. In Figure \ref{fig:samplesegblackswan}, the green dots (located immediately to the right side of the orange dots) indicate the samples drawn from this extremely narrow conditional PDF. In using MCMC samplers to draw from such a conditional PDF, the rate of acceptance will evidently be extremely low leading to a large number of repeated samples which contributes to the large CoV values. Clearly, the proposed sampler is unable to alleviate this difficulty, since it has not been designed to deal with the task of drawing from narrow conditional PDFs.

Nevertheless, the construction of the proposed sampler makes it naturally amenable to be used in conjunction with other samplers. In particular, the local transition PDF $p_{\text{local}}(\cdot\mid \cdot)$, which was implemented as the modified Metropolis transition PDF in this study, can in principle correspond to any valid MCMC algorithm. Thus, to address the geometric challenges such as those encountered in the Black Swan problem, a potential way forward is to use the proposed sampler with more advanced local samplers such as the affine invariant ensemble MCMC sampler (which has been shown to perform well in terms of sampling from narrow PDFs and has been explored in the context of subset simulation by Shields \textit{et al.,}~\cite{shields2021subset}). Indeed, the extensive development of advanced MCMC samplers for subset simulation~\cite{santoso2011modified,zuev2011modified,papaioannou2015mcmc,wang2019hamiltonian,shields2021subset,chen2022riemannian} may be leveraged and used along with the proposed sampler. Hence, future work may consider the best ways to couple various types of MCMC samplers together with the proposed method, with the aim to handle a variety of geometric complexities that may arise in structural engineering applications.

\begin{table}[!ht]
    \centering
    \caption{Numerical results for the Black Swan problem with $n_s=1000$. $\bar{P}_F$ and $\bar{\delta}$ are average failure probability and CoV from 1000 independent trials. $\bar{N}_{\text{calls}}$ is the average number of performance function evaluations from 1000 independent trials. $\bar{\Delta}=\bar{\delta}/\bar{N}_{\text{calls}}$.}
    \label{tab:egblackswan}
    \begin{tabular}{ccccc}
    \toprule
    Method & $\bar{P}_F$ & $\bar{N}_{\text{calls}}$ & $\bar{\delta}$ & $\bar{\Delta}$ \\
    \midrule
    S1 & $7.59\times10^{-9}$ & $7389$ & $3.12$ & $268.92$ \\
    S2 & $8.04\times10^{-9}$ & $8177$ & $2.77$ & $250.80$ \\
    S3 & $3.16\times10^{-5}$ & $3094$ & $0.43$ & $33.85$ \\
    S4 & $9.84\times10^{-9}$ & $15468$ & $1.14 (0.80)$ & $142.74$ \\
    \midrule
    \multicolumn{5}{l}{True probability of failure $P_F = 6.52\times10^{-9}$} \\
    \bottomrule
    \end{tabular}
\end{table}

\section{Conclusions}
\label{section: limitations_and_discussions}
The present study focuses on reliability estimation problems comprising (a) several, possibly disconnected, regions of failure, all of which may not contribute equally to failure probability, or (b) discontinuities or sharp changes in performance function values. In the context of subset simulation, these problems require drawing samples from multimodal conditional standard normal PDFs. Inability to adequately explore and sample from such PDFs lead to gross underestimation of the failure probability. As highlighted by the performance of traditional subset simulation for the illustrative examples, as well as past studies~\cite{breitung2019geometry,rashki2021sesc,kinnear2025niching,li2025relaxed}, we see that the usual implementation of the framework struggles severely when the aforementioned challenges are present. In response, we propose a cyclic component-wise variant of Intrepid MCMC which demonstrates more effective exploration of the standard normal space compared to the samplers currently used in the literature within the framework of subset simulation. The key idea is to use two transition PDFs: (a) $p_{\text{local}}\left(\bm{u}^{(k+1)}\mid \check{\bm{u}}^{(k)}\right)$, which focuses on generating samples from the mode that the chain currently resides in, and (b) $p_\text{global}\left(\check{\bm{u}}^{(k)}\mid\bm{u}^{(k)}\right)$, which focuses on exploring and detecting previously undiscovered modes. These transition PDFs are then combined in a cyclic fashion to obtain the Intrepid transition PDF, $p_{\text{CI}}\left(\bm{u}^{(k+1)}\mid \bm{u}^{(k)}\right)$, as defined in Eq.~\eqref{eq:Intrepid_mtd}. 

The illustrative examples show that the proposed method is able to more thoroughly explore the standard normal space and produce better estimates of failure probability with lower CoV and unit CoV values for a wider variety of problems. Furthermore, for a structural engineering application where performance functions are defined implicitly via black-box computational codes, the analyst would not know a priori whether such geometric difficulties exist. Therefore, when an algorithm produces a failure probability estimate, the analyst would have no means to judge whether the probability of failure was underestimated due to the algorithm missing a failure region. In such settings, the proposed sampler more robustly yields stabler and more accurate estimates of the failure probability by accounting for all important failure regions. 
We emphasize, however, that although the proposed sampler detected the important failure regions in all examples considered here, including in a high-dimensional space (Section \ref{subsection:ex4}), there is no finite-sample guarantee that all such regions will be detected in every high-dimensional problem. However, as with MCMC samplers more generally, sufficiently long runs are guaranteed to eventually explore all regions since the proposed sampler is Harris ergodic (shown in Appendix~\ref{subsection:proof}).
Overall, as demonstrated in the illustrative examples, the proposed method stands a better chance compared to existing MCMC samplers to contend with these geometric difficulties even in high dimensions.

\begin{appendices}
\section{Appendix A: Mathematical validity of the cyclic component-wise Intrepid MCMC sampler}
\label{subsection:proof}

In this section, we show that the proposed cyclic component-wise Intrepid MCMC sampler draws valid samples from the target PDF $\phi(\bm{u}\mid F_l)$. We do so by first showing that the target PDF is stationary with respect to the Intrepid Markov transition PDF $p_{\text{CI}}\left(\bm{u}^{(k+1)}\mid \bm{u}^{(k)}\right)$. Mathematically, we wish to prove that,
\begin{equation}
    \label{eq:global_balance}
    \phi\left(\bm{u}^{(k+1)}\mid F_l\right)=\int{p_{\text{CI}}\left(\bm{u}^{(k+1)}\mid \bm{u}^{(k)}\right)\phi\left(\bm{u}^{(k)}\mid F_l\right)\,d\bm{u}^{(k)}}
\end{equation}
This would imply that if the current state $\bm{u}^{(k)}$ is distributed according to $\phi(\bm{u}\mid F_l)$, then the subsequent state $\bm{u}^{(k+1)}$ is also distributed according to $\phi(\bm{u}\mid F_l)$ and hence the transition PDF is stationary with respect to $p_{\text{CI}}\left(\bm{u}^{(k+1)}\mid \bm{u}^{(k)}\right)$. We begin with the RHS.
\begin{align}
    \label{eq:gb_simple}
    \int{p_{\text{CI}}\left(\bm{u}^{(k+1)}\mid \bm{u}^{(k)}\right)\phi\left(\bm{u}^{(k)}\mid F_l\right)\,d\bm{u}^{(k)}} 
    &= 
    \int{\left(\int{p_{\text{local}}\left(\bm{u}^{(k+1)}\mid \check{\bm{u}}^{(k)}\right)p_\text{global}\left(\check{\bm{u}}^{(k)}\mid\bm{u}^{(k)}\right)}\, d\check{\bm{u}}^{(k)}\right) \,\phi\left(\bm{u}^{(k)}\mid F_l\right)\,d\bm{u}^{(k)}} \nonumber \\
    &=
    \int{\int{p_{\text{local}}\left(\bm{u}^{(k+1)}\mid \check{\bm{u}}^{(k)}\right)p_\text{global}\left(\check{\bm{u}}^{(k)}\mid\bm{u}^{(k)}\right)}\,\phi\left(\bm{u}^{(k)}\mid F_l\right) \,d\bm{u}^{(k)} \, d\check{\bm{u}}^{(k)}} \nonumber \\
    &=\int{p_{\text{local}}\left(\bm{u}^{(k+1)}\mid \check{\bm{u}}^{(k)}\right) \left(\int{p_\text{global}\left(\check{\bm{u}}^{(k)}\mid\bm{u}^{(k)}\right)}\,\phi\left(\bm{u}^{(k)}\mid F_l\right) \,d\bm{u}^{(k)} \right) \, d\check{\bm{u}}^{(k)}}
\end{align}
We prove Eq.~\eqref{eq:global_balance} in two stages: 
\begin{enumerate}[label=\alph*)]
    \item First, we show that the inner integral of Eq.~\eqref{eq:gb_simple} $\int{p_\text{global}\left(\check{\bm{u}}^{(k)}\mid\bm{u}^{(k)}\right)\,\phi\left(\bm{u}^{(k)}\mid F_l\right) \,d\bm{u}^{(k)}}=\phi\left(\check{\bm{u}}^{(k)}\mid F_l\right)$.
    \item Then, we show that the outer integral of Eq.~\eqref{eq:gb_simple} $\int{p_{\text{local}}\left(\bm{u}^{(k+1)}\mid \check{\bm{u}}^{(k)}\right)\phi\left(\check{\bm{u}}^{(k)}\mid F_l\right)\, d\check{\bm{u}}^{(k)}}=\phi\left(\bm{u}^{(k+1)}\mid F_l\right)$.
\end{enumerate}
We begin with the identity given in (a). Note that $p_\text{global}\left(\check{\bm{u}}^{(k)}\mid\bm{u}^{(k)}\right)$ is a component-wise MH transition PDF executed in the hyperspherical coordinate system where all component-wise transitions are independent of one another. Hence, the Markov transition PDF between two states $\bm{u}^{(k)}$ and $\check{\bm{u}}^{(k)}$ (represented by $\bm{u}^{\text{hyp},(k)}$ and $\check{\bm{u}}^{\text{hyp},(k)}$, respectively in the hyperspherical coordinate system) lying in $F_l$ can be expressed as follows.
\begin{equation}
    \label{eq:hyperspherical}
    p_\text{global}^\text{hyp}\left(\check{\bm{u}}^{\text{hyp},(k)}\mid \bm{u}^{\text{hyp},(k)}\right)=\prod_{i=1}^{n_d-1}p_{\Theta_{i}}\left(\check{\theta}_{i}^{(k)}\mid \theta_{i}^{(k)}\right)\,p_R\left(\check r^{(k)} \mid r^{(k)}\right)
\end{equation}
Here, $p_R\left(\check r^{(k)} \mid r^{(k)}\right)$ is the MH transition PDF of the radial component and $p_{\Theta_{i}}\left(\check{\theta}_{i}^{(k)}\mid \theta_{i}^{(k)}\right),i=1,2,\cdots,n_d-1$ represent the MH transition PDF of the angular components. 

\begin{equation}
\begin{aligned}
p_R\left(\check r \mid r^{(k)}\right)
&= q_R\left(\check r^{(k)} \mid r^{(k)}\right)\,\alpha_R\left(r^{(k)},\check r^{(k)}\right) \\
&\quad + \left(
1 - \int q_R\left(z \mid r^{(k)}\right)\,\alpha_R\left(r^{(k)},z\right)\,dz
\right)\delta\left(\check r^{(k)} - r^{(k)}\right),
\\[6pt]
p_{\Theta_{i}}\left(\check \theta_{i}^{(k)} \mid \theta_{i}^{(k)}\right)
&= q_{\Theta_{i}}\left(\check \theta_{i}^{(k)} \mid \theta_{i}^{(k)}\right)
   \,\alpha_{\Theta_{i}}\left(\theta_{i}^{(k)},\check \theta_{i}^{(k)}\right) \\
&\quad + \left(
1 - \int q_{\Theta_{i}}\left(z \mid \theta_{i}^{(k)}\right)\,\alpha_{\Theta_{i}}\left(\theta_{i}^{(k)},z\right)\,dz
\right)\delta\left(\check \theta_{i}^{(k)} - \theta_{i}^{(k)}\right),
\\
&\qquad \text{where } i = 1,2,\dots,n_d-1.
\end{aligned}
\end{equation}

Therefore, they satisfy the detailed balance condition with respect to their respective target PDFs. This leads to the following set of equations.
\begin{align}
    p_R\left(\check r^{(k)} \mid r^{(k)}\right)f_R\left(r^{(k)}\right)&=p_R\left(r^{(k)} \mid \check r^{(k)}\right)f_R\left(\check r^{(k)}\right) \label{eq:radial_db}\\
    p_{\Theta_{i}}\left(\check \theta_{i}^{(k)} \mid \theta_{i}^{(k)}\right)f_{\Theta_{i}}\left(\theta_{i}^{(k)}\right)&=p_{\Theta_{i}}\left(\theta_{i}^{(k)} \mid \check \theta_{i}^{(k)}\right)f_{\Theta_{i}}\left(\check \theta_{i}^{(k)}\right), \text{for } i=1,2,\cdots,n_d-1 \label{eq:angular_db}
\end{align}
Multiplying Eq.~\eqref{eq:radial_db} and Eq.~\eqref{eq:angular_db} yields the following identity.
\begin{equation}
        \label{eq:complete_db}
        p_R\left(\check r \mid r^{(k)}\right)f_R\left(r^{(k)}\right) \prod_{i=1}^{n_d-1}{p_{\Theta_{i}}\left(\check \theta_{i} \mid \theta_{i}^{(k)}\right)}f_{\Theta_{i}}\left(\check \theta_{i}\right)=p_R\left(r^{(k)} \mid \check r\right)f_R(\check r) \prod_{i=1}^{n_d-1}p_{\Theta_{i}}\left(\theta_{i}^{(k)}\mid \check \theta_{i}\right)f_{\Theta_{i}}(\check \theta_{i})\\
\end{equation}
Using Eq.~\eqref{eq:complete_db} and Eq.~\eqref{eq:hyperspherical}, we get,
\begin{equation}
    \begin{split}
    p_\text{global}^\text{hyp}\left(\check{\bm{u}}^{\text{hyp},(k)}\mid\bm{u}^{\text{hyp},(k)}\right)f_{\bm{U}^\text{hyp}}\left(\bm{u}^{\text{hyp},(k)}\right)
    &=p_\text{global}^\text{hyp}\left(\bm{u}^{\text{hyp},(k)}\mid \check{\bm{u}}^{\text{hyp},(k)}\right)f_{\bm{U}^\text{hyp}}\left(\check{\bm{u}}^{\text{hyp},(k)}\right)\\
    \end{split}
\end{equation}
Transforming back to the Cartesian coordinates, we have $\check{\bm{u}}^{\text{hyp},(k)}=\Psi\left(\check{\bm{u}}^{(k)}\right)$ and $\bm{u}^{\text{hyp},(k)}=\Psi\left(\bm{u}^{(k)}\right)$ which gives,

\begin{align}
    &p_\text{global}^\text{hyp}\left(\Psi\left(\check{\bm{u}}^{(k)}\right)\mid\Psi\left(\bm{u}^{(k)}\right)\right)
    f_{\bm{U}^\text{hyp}}\left(\Psi\left(\bm{u}^{(k)}\right)\right) \nonumber \\
    &\qquad = p_\text{global}^\text{hyp}\left(\Psi\left(\bm{u}^{(k)}\right)\mid \Psi\left(\check{\bm{u}}^{(k)}\right)\right)
    f_{\bm{U}^\text{hyp}}\left(\Psi\left(\check{\bm{u}}^{(k)}\right)\right) \\[1em]
    &\Rightarrow p_\text{global}^\text{hyp}\left(\Psi\left(\check{\bm{u}}^{(k)}\right)\mid\Psi\left(\bm{u}^{(k)}\right)\right)
    \left|\det\left(J^{\bm{u}^\text{hyp}}_{\bm{u}}\right)\right|
    f_{\bm{U}^\text{hyp}}(\Psi(\bm{u}^{(k)}))
    \left|\det\left(J^{\bm{u}^\text{hyp}}_{\bm{u}}\right)\right| \nonumber \\
    &\qquad = p_\text{global}^\text{hyp}\left(\Psi\left(\bm{u}^{(k)}\right)\mid \Psi\left(\check{\bm{u}}^{(k)}\right)\right)
    \left|\det\left(J^{\bm{u}^\text{hyp}}_{\bm{u}}\right)\right|
    f_{\bm{U}^\text{hyp}}\left(\Psi\left(\check{\bm{u}}^{(k)}\right)\right)
    \left|\det\left(J^{\bm{u}^\text{hyp}}_{\bm{u}}\right)\right| \\[1em]
    &\Rightarrow p_\text{global}\left(\check{\bm{u}}^{(k)}\mid\bm{u}^{(k)}\right)\phi\left(\bm{u}^{(k)}\right)
    = p_\text{global}\left(\bm{u}^{(k)}\mid \check{\bm{u}}^{(k)}\right)\phi\left(\check{\bm{u}}^{(k)}\right)
\end{align}

Since both $\bm{u}^{(k)}$ and $\check{\bm{u}}^{(k)}$ must lie in the failure region, this implies,
\begin{equation}
    \label{eq:global_db}
    \begin{split}
        p_\text{global}\left(\check{\bm{u}}^{(k)}\mid\bm{u}^{(k)}\right)\phi\left(\bm{u}^{(k)}\mid F_l\right)
        =p_\text{global}\left(\bm{u}^{(k)}\mid \check{\bm{u}}^{(k)}\right)\phi\left(\check{\bm{u}}^{(k)}\mid F_l\right)
    \end{split}
\end{equation}
This allows us to prove the identity stated in (a). The LHS of Eq.~\eqref{eq:gb_simple} reads as follows.
\begin{align}
\int{p_\text{global}\left(\check{\bm{u}}^{(k)}\mid\bm{u}^{(k)}\right)\,\phi\left(\bm{u}^{(k)}\mid F_l\right) \,d\bm{u}^{(k)}} &= \int{p_\text{global}\left(\bm{u}^{(k)}\mid \check{\bm{u}}^{(k)}\right)\phi\left(\check{\bm{u}}^{(k)}\mid F_l\right) \,d\bm{u}^{(k)}} \text{ (from Eq.~\eqref{eq:global_db})}\\
&=\phi\left(\check{\bm{u}}^{(k)}\mid F_l\right)\int{p_\text{global}\left(\bm{u}^{(k)}\mid \check{\bm{u}}^{(k)}\right) \,d\bm{u}^{(k)}}\\
&=\phi\left(\check{\bm{u}}^{(k)}\mid F_l\right), \text{ as desired.}
\end{align}
Now we show the identity mentioned in (b). $p_\text{local}(\bm{u}^{(k+1)}\mid\check{\bm{u}}^{(k)})$ is a component-wise MH transition PDF executed in the Cartesian coordinate system, where each component-wise transition is independent. Therefore, the transition PDF between states $\check{\bm{u}}^{(k)}$ and $\bm{u}^{(k+1)}$ is given by,
\begin{equation}
    \label{eq:cartesian_decomposition}
    p_\text{local}\left(\bm{u}^{(k+1)}\mid \check{\bm{u}}^{(k)}\right)=\prod_{i=1}^{n_d}p_{U_{i}}\left(u_{i}^{(k)}\mid \check{u}_{i}^{(k)}\right)
\end{equation}
Here, $p_{U_{i}}\left(u_{i}^{(k+1)}\mid \check{u}_{i}^{(k)}\right)$ represent the MH transition PDF for $i^\text{th}$ Cartesian component of $\bm{U}$ which can be written as follows.

\begin{align}
p_{U_{i}}\left(u_{i}^{(k+1)}\mid \check{u}_{i}^{(k)}\right)
&= q_{U_{i}}\left(u_{i}^{(k+1)} \mid \check{u}_{i}^{(k)}\right)
   \,\alpha_{U_{i}}\left(\check{u}_{i}^{(k)}, u_{i}^{(k+1)}\right) \\
&\quad + \left(
1 - \int q_{U_{i}}\left(z \mid \check{u}_{i}^{(k)}\right)
        \,\alpha_U\left(\check{u}_{i}^{(k)}, z\right)\,dz
\right)
\delta\left(u_{i}^{(k+1)} - \check{u}_{i}^{(k)}\right),
\\
&\qquad \text{where }i = 1,2,\dots,n_d.
\end{align}

Each transition PDF must satisfy the detailed balance condition with respect to its corresponding target PDF. Therefore, we have the following.
\begin{equation}
    \label{eq:cartesian_db}
    p_{U_{i}}\left(u_{i}^{(k+1)} \mid \check u_{i}^{(k)}\right)f_{U_{i}}\left(\check u_{i}^{(k)}\right) = p_{U_{i}}\left(\check u_{i}^{(k)} \mid u_{i}^{(k+1)}\right)f_{U_{i}}\left(u_{i}^{(k+1)}\right), \text{for } i=1,2,\cdots,n_d
\end{equation}
Multiplying Eq.~\eqref{eq:cartesian_decomposition} for $i=1,2,\cdots,n_d$ and using Eq.~\eqref{eq:cartesian_db},
\begin{equation}
    p_\text{local}\left(\bm{u}^{(k+1)}\mid \check{\bm{u}}^{(k)}\right)\phi\left(\check{\bm{u}}^{(k)}\right)
    =p_\text{local}\left(\check{\bm{u}}^{(k)} \mid \bm{u}^{(k+1)}\right)\phi\left(\bm{u}^{(k+1)}\right)
\end{equation}
Since $\check{\bm{u}}^{(k)}$ and $\bm{u}^{(k+1)}$ lie in the failure region $F_l$, we have,
\begin{equation}
    \label{eq:local_db}
    p_\text{local}\left(\bm{u}^{(k+1)}\mid \check{\bm{u}}^{(k)}\right)\phi\left(\check{\bm{u}}^{(k)}\mid F_l\right)
    =p_\text{local}\left(\check{\bm{u}}^{(k)} \mid \bm{u}^{(k+1)}\right)\phi\left(\bm{u}^{(k+1)}\mid F_l\right)
\end{equation}
This allows us to prove the identity stated in (b). The LHS is given by,
\begin{align}
\int{p_{\text{local}}\left(\bm{u}^{(k+1)}\mid \check{\bm{u}}^{(k)}\right)\phi\left(\check{\bm{u}}^{(k)}\mid F_l\right)\, d\check{\bm{u}}^{(k)}} &= \int{p_\text{local}\left(\check{\bm{u}}^{(k)} \mid \bm{u}^{(k+1)}\right)\phi\left(\bm{u}^{(k+1)}\mid F_l\right)}\,d\check{\bm{u}}^{(k)} \text{ (from Eq.~\eqref{eq:local_db})}\\
&=\phi\left(\bm{u}^{(k+1)}\mid F_l\right)\int{p_\text{local}\left(\check{\bm{u}}^{(k)} \mid \bm{u}^{(k+1)}\right)}\,d\check{\bm{u}}^{(k)}\\
&=\phi\left(\bm{u}^{(k+1)}\mid F_l\right), \text{as desired.}
\end{align}
We have now proven the identities mentioned in (a) and (b). This implies that Eq.~\eqref{eq:global_balance} must be true and that the proposed Intrepid MCMC transition PDF is indeed stationary with respect to the target PDF $\phi(\bm{u} \mid F_l)$. Furthermore, we note that $p_{\text{Intrepid}}(\bm{u}'\mid \bm{u})>0$ for all $\bm{u}',\bm{u}\in F_l$ which implies that the Markov chain is Harris ergodic~\cite{brooks2011handbook}. Having proven the chain is Harris ergodic, we remark on the following two properties of the chain.

\begin{enumerate}
    \item \textbf{Strong law of large numbers for Markov chains}: Let us denote the true conditional probability to be estimated as $P_l=\mathbb{P}(F_{l+1}\mid F_l)$. For Harris ergodic chains, if $\bm{u}^{(1)},\bm{u}^{(2)},\cdots,\bm{u}^{(n_s)}$ are samples generated from a single Markov chain with stationary PDF $\phi(\bm{u} \mid F_l)$, the following holds~\cite{brooks2011handbook}.

    \begin{equation}
        \label{eq:slln}
        \hat{P}_l=\frac{1}{n_s}\sum_{i=1}^{n_s} I\left[G(\bm{u}^{(i)})\leq b_l\right] \xrightarrow{\text{a.s.}} P_l \text{ as } n_s\rightarrow \infty
    \end{equation}
    
    Therefore, we rely on Eq.~\eqref{eq:slln} to estimate intermediate thresholds $b_l$, $l=2,3,\cdots,n_f$.

    \item \textbf{Central limit theorem for Markov chains}: In order to estimate the CoV of $\hat{P}_l$, we invoke the central limit theorem for Markov chains. The approximate sampling variance of $\hat{P}_l$ is given by the Markov chain central limit theorem if there exists a constant $\sigma_l^2$ such that as $n_s\rightarrow\infty$, the following holds~\cite{Flegal2010}.

    \begin{equation}
        \label{eq:clt}
        \sqrt{n_s}\left(\hat{P}_l-P_l  \right)\xrightarrow{d} \text{N}(0,\sigma_l^2)
    \end{equation}
    
    If a Markov chain is Harris ergodic and reversible, the existence of such a constant $\sigma_l^2$ is guaranteed~\cite{geyer1992practical}. The proposed sampler has been shown to be Harris ergodic in Section \ref{subsection:proof}, but it is not reversible. Absent the property of reversibility, mathematically establishing the existence of $\sigma_l^2$ requires proving regularity conditions relating to, for example, mixing coefficients of the Markov chain, drift conditions, or conditions of geometric, uniform or polynomial ergodicity, etc. A detailed discussion on these conditions can be found in~\cite{Jones2004}. However, examining these regularity conditions for the proposed non-reversible sampler is technically involved and therefore not pursued in the present study. Instead, here, we show empirically that for a particular example, when employing the cyclic component-wise Intrepid sampler, the quantity $\sqrt{n_s}\left(\hat{P}_l-P_l\right)$ is distributed as $\text{N}(0,\sigma_l^2)$ for some constant $\sigma_l^2$.

    In order to empirically verify this, we consider a simple the performance function defined as follows.
    \begin{equation}
        G(\bm{U})=-U_1-U_2
    \end{equation}
    
    We define $P_l=\mathbb{P}\left(G(\bm{U})\leq 0\right)$. It is evident that the true value $P_l=0.5$. We initiate a Markov chain using the component-wise Intrepid sampler with target PDF $\phi(\bm{u}\mid F_l)$, where $F_l=\{\bm{u} \in \mathbb{R}^2: G(\bm{u})\leq 0\}$. Let $\bm{u}^{(1)}, \bm{u}^{(2)},\cdots,\bm{u}^{(n_s)}$ be the Markov chain samples with $\bm{u}_1=(0.1, 0.1)$. Accordingly, the estimator $\hat P_l$ is defined as follows.
    \begin{equation}
        \hat P_l=\frac{1}{n_s}\sum_{i=1}^{n_s} I\left[G(\bm{u}^{(i)})\leq 0\right]
    \end{equation}
    
    1000 independent realizations of the estimator $\hat P_l$ are generated from which empirical CDFs of $\sqrt{n_s}\left(\hat{P}_l-P_l \right)$ are plotted for $n_s=10,50,100,200,500,1000,$ and $10000$ samples as shown in Figure \ref{fig:clt_cdf}. The figure also contains the CDF of the normal distribution $\text{N}(0,\hat \sigma_{n_s}^2)$, where $\hat \sigma_{n_s}^2$ is the empirical variance obtained from the realizations of $\hat P_l$. Figure \ref{fig:clt_cdf} and Table \ref{tab:clt_verification} show the KS-statistic between the empirical CDF and the normal CDF. The figure as well as the numerical values of the KS-statistic indicate that KS-statistic between the two CDFs decreases as $n_s$ increases. Table \ref{tab:clt_verification} also shows the values of $\hat\sigma_{n_s}$ as $n_s$ increases. Therefore, the quantity $\sqrt{n_s}\left(\hat{P}_l-P_l \right)$ is empirically shown to converge in distribution to $\text{N}(0,\hat \sigma_{n_s}^2)$.
    
    \begin{figure}[htbp]
    \centering
    \includegraphics[scale=0.45]{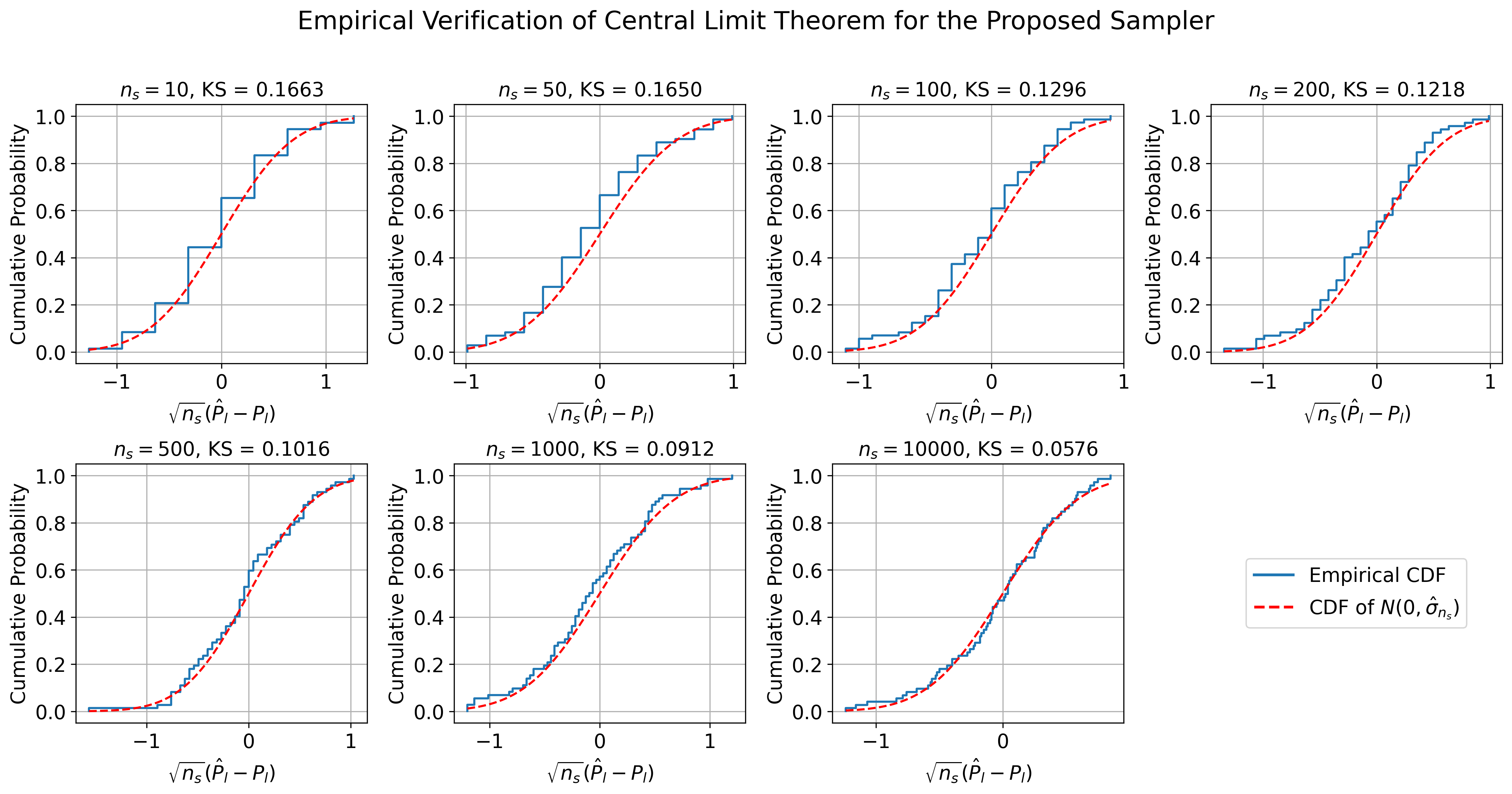}  
    \caption{Comparison of empirical CDF of $\sqrt{n_s}\left(\hat{P}_l-P_l \right)$ and $\text{N}(0,\hat \sigma_{n_s}^2)$.}
    \label{fig:clt_cdf}
    \end{figure}

    \begin{table}[htbp]
        \centering
        \caption{KS-statistic and $\hat\sigma_{n_s}$ for various values of $n_s$}
        \label{tab:clt_verification}
        \begin{tabular}{ccc}
            \hline
            $n_s$ & KS-statistic & $\hat\sigma_{n_s}$ \\
            \hline
            10 & 0.16 & 0.53 \\
            50 & 0.16 & 0.45 \\
            100 & 0.12 & 0.43 \\
            200 & 0.12 & 0.48 \\
            500 & 0.10 & 0.50 \\
            1000 & 0.09 & 0.53 \\
            10000 & 0.05 & 0.46 \\
            \hline
        \end{tabular}
    \end{table}

\end{enumerate}

\section{Appendix B: Derivation of load-deflection relationship and total potential energy of the von-Mises truss}
\label{subsection:load_def_derivation}

\begin{figure}[htbp]
\centering
\includegraphics[scale=0.5]{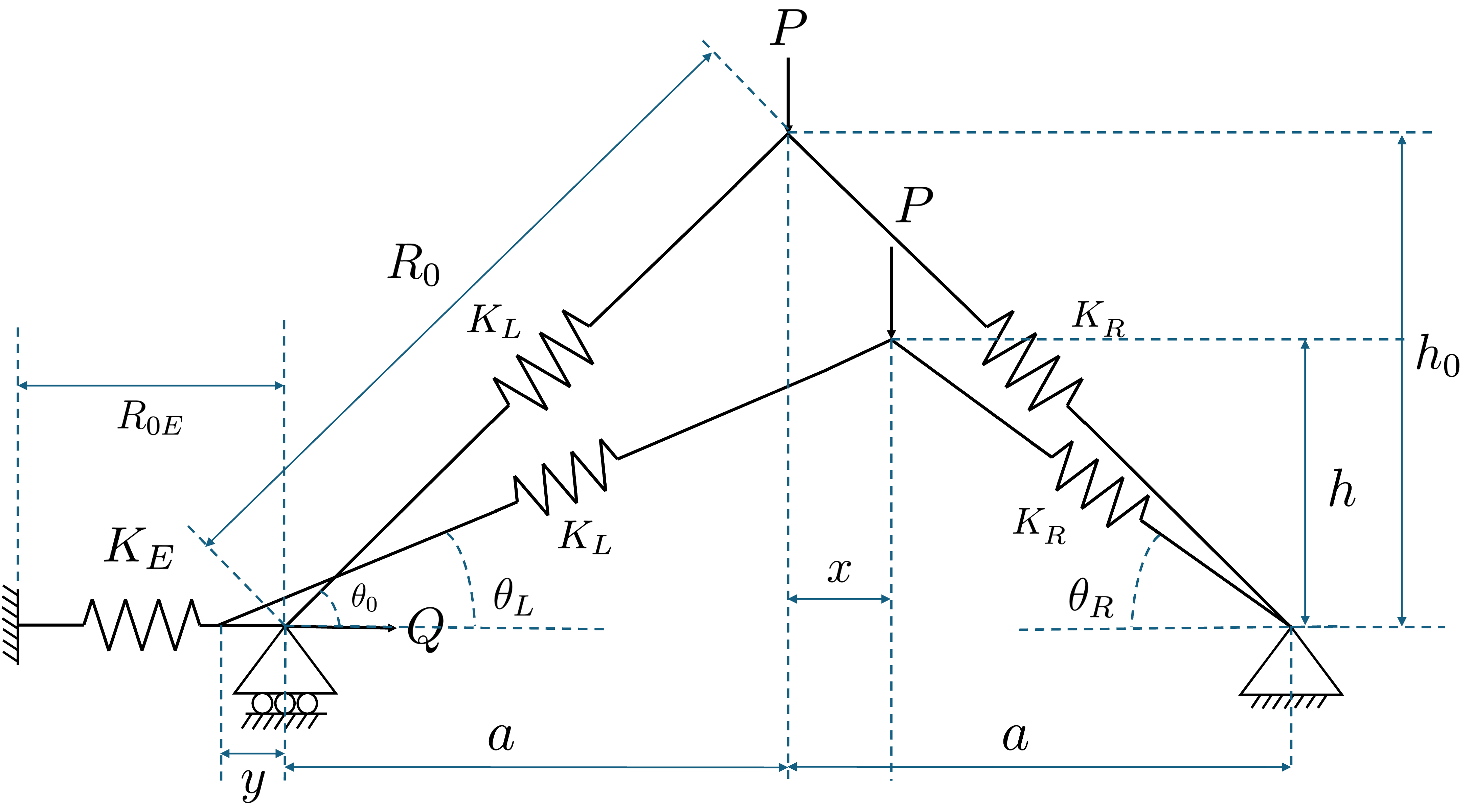}  
\caption{Undeformed and deformed configurations of the von-Mises truss.}
\label{fig:def_undef_ex4}
\end{figure}

In this section, we derive the equation for the load-deflection relationship between $P$ and $\theta_L$, and the total potential energy associated with the von-Mises truss considered in Section~\ref{subsection:ex4}. Figure~\ref{fig:def_undef_ex4} depicts the undeformed and deformed configuration of the von-Mises truss. 

As described in Section~\ref{subsection:ex4}, the top vertex is subjected to a downward vertical load $P$ and the spring with stiffness $K_E$ is subjected to a horizontal load $Q$. $K_L$ is the resultant stiffness arising from 500 springs of stiffnesses $K_L^1, K_L^2, \cdots, K_L^{500}$ arranged in series, and $K_R$ is the resultant stiffness arising from 500 springs of stiffnesses $K_R^1, K_R^2, \cdots, K_R^{500}$ arranged in series.

In the undeformed configuration, $R_0$ denotes the arm length which are equal on both sides, $R_{0E}$ denotes the length of the spring attached to the roller, $2a$ is the total horizontal length, $h_0$ is the total height of the truss, and $\theta_0$ is the angle made by the left and right arms with respect to the horizontal. Upon deforming, the left and the right arm lengths change to $R_L$ and $R_R$, respectively. The length spring attached to the roller becomes $R_E$. The angle made by the left and the right arms with respect to the horizontal change to $\theta_L$ and $\theta_R$, respectively. The rightward horizontal distance covered by the top vertex of the truss is denoted by $x$, and the leftward distance traveled by the roller support is $y$. The height of the truss upon deforming is $h$. 

The following geometric relationships hold.
\begin{align}
    \label{eqn:geometric_relationship_1}
    a+x+y = R_L \cos{\theta_L}\\
    \label{eqn:geometric_relationship_2}
    a-x = R_R \cos{\theta_R}
\end{align}

Eq.~\eqref{eqn:geometric_relationship_1} and Eq.~\eqref{eqn:geometric_relationship_2} leads to

\begin{align}
    \label{eqn:expression_y}
    y=h\cot{\theta_L}+h\cot{\theta_R}-2a\\
\end{align}

Forces experience by springs $K_L$, $K_R$, and $K_E$ are, respectively:
\begin{align}
    \label{eqn:force_v_top}
    F_L &= -K_L(R_L-R_0)=K_L\left(R_0-\frac{h}{\sin{\theta_L}}\right)\\
    \label{eqn:force_h_top}
    F_R &= -K_R(R_R-R_0)=K_R\left(R_0-\frac{h}{\sin{\theta_R}}\right)\\
    \label{eqn:force_h_roller}
    F_E &= -K_E(R_E-R_{0E})=K_Ey
\end{align}

Together with Eq.~\eqref{eqn:force_v_top}, Eq.~\eqref{eqn:force_h_top}, and Eq.~\eqref{eqn:force_h_roller}, we write the following equilibrium equations. 

Firstly, the horizontal equilibrium equation at the roller support takes the following form.

\begin{equation}
    \label{eqn:eq_h_roller}
    Q+k_Ey = K_L\left(R_0-\frac{h}{\sin{\theta_L}}\right)\cos{\theta_L}\\
\end{equation}
    
Substituting Eq.~\eqref{eqn:expression_y} into Eq.~\eqref{eqn:eq_h_roller} yields,
\begin{align}
    &Q+ K_E\left[h\left(\cot\theta_L+\cot\theta_R\right)-2a\right] = K_L\left(R_0\cos\theta_L-h\cot\theta_L\right)\\
    &\Rightarrow h\left[ K_E\left(\cot\theta_L+\cot\theta_R\right) +K_L\cot\theta_L \right] = 2aK_E+K_LR_0\cos\theta_L-Q\\
    \label{eqn:expression_h}
    &\Rightarrow h = \frac{2aK_E+K_LR_0\cos\theta_L-Q}{ K_E\left(\cot\theta_L+\cot\theta_R\right) +K_L\cot\theta_L}
\end{align}

Secondly, the horizontal equilibrium equation at the top vertex is written as follows.

\begin{align}
    K_L\left(R_0-\frac{h}{\sin{\theta_L}}\right)\cos{\theta_L} &= K_R\left(R_0-\frac{h}{\sin{\theta_R}}\right)\cos{\theta_R}\\
    \label{eqn:theta_l_theta_r}
    \Rightarrow h\left(K_L\cot\theta_L-K_R\cot\theta_R\right) &= R_0\left(K_L\cos\theta_L-K_R\cos\theta_R\right)
\end{align}

Finally, the vertical equilibrium equation at the top vertex is written as follows.

\begin{equation}
    \label{eqn:load_deflection}
    P=K_L\left(R_0-\frac{h}{\sin{\theta_L}}\right)\sin{\theta_L}+K_R\left(R_0-\frac{h}{\sin{\theta_R}}\right)\sin{\theta_R}
\end{equation}

Therefore, the equation that relates the vertical load $P$ to $\theta_L$ is given by Eq.~\eqref{eqn:load_deflection}, with Eq.~\ref{eqn:theta_l_theta_r} relating $\theta_L$ with $\theta_R$, and Eq.~\eqref{eqn:expression_h} providing the expression for $h$.

While Eq.~\eqref{eqn:load_deflection} represents the equilibrium equation, it does not provide any information as to whether a given equilibrium configuration is stable. In order to determine stability, we must consider the total potential energy of the system which can be written as follows.

\begin{equation}
    \Pi=\frac{1}{2}K_L\left( R_0-\frac{h}{\sin{\theta_L}} \right)^2 + \frac{1}{2}K_R\left( R_0-\frac{h}{\sin{\theta_R}} \right)^2 + \frac{1}{2}k_Ey^2+Qy-P(h_0-h)
\end{equation}

Here, $h$ is given by Eq.~\eqref{eqn:expression_h} and $y$ can be computed by rearranging Eq.~\eqref{eqn:eq_h_roller} as $y=\left[K_L\left( R_0\cos{\theta_L}-h\cot{\theta_L}\right) - Q\right]/{K_E}$. Stability is determined by numerically computing the second derivative of $\Pi$ with respect to $\theta_L$.

\end{appendices}

\bibliographystyle{plain}
\bibliography{references}

@book{melchers2018structural,
  title={Structural reliability analysis and prediction},
  author={Melchers, Robert E and Beck, Andr{\'e} T},
  year={2018},
  publisher={John wiley \& sons}
}

@article{hasofer1974exact,
  title={Exact and invariant second-moment code format},
  author={Hasofer, Abraham M and Lind, Niels C},
  journal={Journal of the Engineering Mechanics division},
  volume={100},
  number={1},
  pages={111--121},
  year={1974},
  publisher={American Society of Civil Engineers}
}

@article{breitung1984asymptotic,
  title={Asymptotic approximations for multinormal integrals},
  author={Breitung, Karl},
  journal={Journal of engineering mechanics},
  volume={110},
  number={3},
  pages={357--366},
  year={1984},
  publisher={American Society of Civil Engineers}
}

@article{rackwitz2001reliability,
  title={Reliability analysis—a review and some perspectives},
  author={Rackwitz, R{\"u}diger},
  journal={Structural safety},
  volume={23},
  number={4},
  pages={365--395},
  year={2001},
  publisher={Elsevier}
}

@article{hohenbichler1988improvement,
  title={Improvement of second-order reliability estimates by importance sampling},
  author={Hohenbichler, Michael and Rackwitz, Ruediger},
  journal={Journal of Engineering Mechanics},
  volume={114},
  number={12},
  pages={2195--2199},
  year={1988},
  publisher={American Society of Civil Engineers}
}

@book{au2014engineering,
  title={Engineering risk assessment with subset simulation},
  author={Au, Siu-Kui and Wang, Yu},
  year={2014},
  publisher={John Wiley \& Sons}
}

@article{tabandeh2022review,
  title={A review and assessment of importance sampling methods for reliability analysis},
  author={Tabandeh, Armin and Jia, Gaofeng and Gardoni, Paolo},
  journal={Structural Safety},
  volume={97},
  pages={102216},
  year={2022},
  publisher={Elsevier}
}

@article{song2023monte,
  title={Monte Carlo and variance reduction methods for structural reliability analysis: A comprehensive review},
  author={Song, Chenxiao and Kawai, Reiichiro},
  journal={Probabilistic Engineering Mechanics},
  volume={73},
  pages={103479},
  year={2023},
  publisher={Elsevier}
}

@article{moustapha2022active,
  title={Active learning for structural reliability: Survey, general framework and benchmark},
  author={Moustapha, Maliki and Marelli, Stefano and Sudret, Bruno},
  journal={Structural Safety},
  volume={96},
  pages={102174},
  year={2022},
  publisher={Elsevier}
}

@article{afshari2023deep,
  title={Deep learning-based methods in structural reliability analysis: a review},
  author={Afshari, Sajad Saraygord and Zhao, Chuan and Zhuang, Xinchen and Liang, Xihui},
  journal={Measurement Science and Technology},
  volume={34},
  number={7},
  pages={072001},
  year={2023},
  publisher={IOP Publishing}
}

@article{breitung2019geometry,
  title={The geometry of limit state function graphs and subset simulation: Counterexamples},
  author={Breitung, Karl},
  journal={Reliability Engineering \& System Safety},
  volume={182},
  pages={98--106},
  year={2019},
  publisher={Elsevier}
}

@article{rashki2021sesc,
  title={SESC: A new subset simulation method for rare-events estimation},
  author={Rashki, Mohsen},
  journal={Mechanical Systems and Signal Processing},
  volume={150},
  pages={107139},
  year={2021},
  publisher={Elsevier}
}

@article{sharma2023modified,
  title={Modified replica exchange-based MCMC algorithm for estimation of structural reliability based on particle splitting method},
  author={Sharma, Adwait and Manohar, CS},
  journal={Probabilistic Engineering Mechanics},
  volume={72},
  pages={103448},
  year={2023},
  publisher={Elsevier}
}

@article{kinnear2025niching,
  title={Niching subset simulation},
  author={Kinnear, Hugh J and DiazDelaO, FA},
  journal={Probabilistic Engineering Mechanics},
  volume={79},
  pages={103729},
  year={2025},
  publisher={Elsevier}
}

@article{au2001estimation,
  title={Estimation of small failure probabilities in high dimensions by subset simulation},
  author={Au, Siu-Kui and Beck, James L},
  journal={Probabilistic engineering mechanics},
  volume={16},
  number={4},
  pages={263--277},
  year={2001},
  publisher={Elsevier}
}

@article{ching2005reliability,
  title={Reliability estimation for dynamical systems subject to stochastic excitation using subset simulation with splitting},
  author={Ching, Jianye and Au, Siu-Kui and Beck, James L},
  journal={Computer methods in applied mechanics and engineering},
  volume={194},
  number={12-16},
  pages={1557--1579},
  year={2005},
  publisher={Elsevier}
}

@article{katafygiotis2007application,
  title={Application of spherical subset simulation method and auxiliary domain method on a benchmark reliability study},
  author={Katafygiotis, LS and Cheung, SH},
  journal={Structural Safety},
  volume={29},
  number={3},
  pages={194--207},
  year={2007},
  publisher={Elsevier}
}

@article{cui2019implementation,
  title={Implementation of machine learning techniques into the subset simulation method},
  author={Cui, Fengkun and Ghosn, Michel},
  journal={Structural Safety},
  volume={79},
  pages={12--25},
  year={2019},
  publisher={Elsevier}
}

@article{zuev2012bayesian,
  title={Bayesian post-processor and other enhancements of Subset Simulation for estimating failure probabilities in high dimensions},
  author={Zuev, Konstantin M and Beck, James L and Au, Siu-Kui and Katafygiotis, Lambros S},
  journal={Computers \& structures},
  volume={92},
  pages={283--296},
  year={2012},
  publisher={Elsevier}
}

@article{santoso2011modified,
  title={Modified Metropolis--Hastings algorithm with reduced chain correlation for efficient subset simulation},
  author={Santoso, AM and Phoon, KK and Quek, ST},
  journal={Probabilistic Engineering Mechanics},
  volume={26},
  number={2},
  pages={331--341},
  year={2011},
  publisher={Elsevier}
}

@article{zuev2011modified,
  title={Modified Metropolis--Hastings algorithm with delayed rejection},
  author={Zuev, Konstantin M and Katafygiotis, Lambros S},
  journal={Probabilistic Engineering Mechanics},
  volume={26},
  number={3},
  pages={405--412},
  year={2011},
  publisher={Elsevier}
}

@article{papaioannou2015mcmc,
  title={MCMC algorithms for subset simulation},
  author={Papaioannou, Iason and Betz, Wolfgang and Zwirglmaier, Kilian and Straub, Daniel},
  journal={Probabilistic Engineering Mechanics},
  volume={41},
  pages={89--103},
  year={2015},
  publisher={Elsevier}
}

@article{wang2019hamiltonian,
  title={Hamiltonian Monte Carlo methods for subset simulation in reliability analysis},
  author={Wang, Ziqi and Broccardo, Marco and Song, Junho},
  journal={Structural Safety},
  volume={76},
  pages={51--67},
  year={2019},
  publisher={Elsevier}
}

@article{chen2022riemannian,
  title={Riemannian Manifold Hamiltonian Monte Carlo based subset simulation for reliability analysis in non-Gaussian space},
  author={Chen, Weiming and Wang, Ziqi and Broccardo, Marco and Song, Junho},
  journal={Structural Safety},
  volume={94},
  pages={102134},
  year={2022},
  publisher={Elsevier}
}

@article{shields2021subset,
  title={Subset simulation for problems with strongly non-Gaussian, highly anisotropic, and degenerate distributions},
  author={Shields, Michael D and Giovanis, Dimitris G and Sundar, VS},
  journal={Computers \& Structures},
  volume={245},
  pages={106431},
  year={2021},
  publisher={Elsevier}
}

@article{chakroborty2026intrepid,
  title={Intrepid MCMC: Metropolis-Hastings with exploration},
  author={Chakroborty, Promit and Shields, Michael D},
  journal={Computer Methods in Applied Mechanics and Engineering},
  volume={448},
  pages={118402},
  year={2026},
  publisher={Elsevier}
}

@article{tierney1994markov,
  title={Markov chains for exploring posterior distributions},
  author={Tierney, Luke},
  journal={the Annals of Statistics},
  pages={1701--1728},
  year={1994},
  publisher={JSTOR}
}

@article{Jones2004,
  title = {On the Markov chain central limit theorem},
  volume = {1},
  ISSN = {1549-5787},
  url = {http://dx.doi.org/10.1214/154957804100000051},
  DOI = {10.1214/154957804100000051},
  number = {none},
  journal = {Probability Surveys},
  publisher = {Institute of Mathematical Statistics},
  author = {Jones,  Galin L.},
  year = {2004},
  month = jan 
}

@book{brooks2011handbook,
  title={Handbook of markov chain monte carlo},
  author={Brooks, Steve and Gelman, Andrew and Jones, Galin and Meng, Xiao-Li},
  year={2011},
  publisher={CRC press}
}

@article{Flegal2010,
  title = {Batch means and spectral variance estimators in Markov chain Monte Carlo},
  volume = {38},
  ISSN = {0090-5364},
  url = {http://dx.doi.org/10.1214/09-AOS735},
  DOI = {10.1214/09-aos735},
  number = {2},
  journal = {The Annals of Statistics},
  publisher = {Institute of Mathematical Statistics},
  author = {Flegal,  James M. and Jones,  Galin L.},
  year = {2010},
  month = apr 
}

@article{geyer1992practical,
  title={Practical markov chain monte carlo},
  author={Geyer, Charles J},
  journal={Statistical science},
  pages={473--483},
  year={1992},
  publisher={JSTOR}
}

@article{gupta2025estimating,
  title={Estimating Monte Carlo variance from multiple Markov chains},
  author={Gupta, Kushagra and Vats, Dootika},
  journal={Scandinavian Journal of Statistics},
  year={2025},
  publisher={Wiley Online Library}
}

@article{au2007application,
  title={Application of subset simulation methods to reliability benchmark problems},
  author={Au, Siu Kui and Ching, Jianye and Beck, James L},
  journal={Structural safety},
  volume={29},
  number={3},
  pages={183--193},
  year={2007},
  publisher={Elsevier}
}

@article{schueller2007benchmark,
  title={Benchmark study on reliability estimation in higher dimensions of structural systems--an overview},
  author={Schu{\"e}ller, Gerhart I and Pradlwarter, Helmuth J},
  journal={Structural Safety},
  volume={29},
  number={3},
  pages={167--182},
  year={2007},
  publisher={Elsevier}
}

@article{chakroborty2025tail,
  title={The tail wags the distribution: Only sample the tails for efficient reliability analysis},
  author={Chakroborty, Promit and Shields, Michael D},
  journal={arXiv preprint arXiv:2505.18510},
  year={2025}
}

@inproceedings{sharma2023sampling,
  title={Sampling Variance Reduction in Structural Reliability Estimation via Sequential MCMC Sampling},
  author={Sharma, Adwait and Manohar, CS},
  booktitle={Advances in Reliability and Safety Assessment for Critical Systems: Proceedings of the 5th National Conference on Reliability and Safety (NCRS 2022)},
  pages={243},
  year={2023},
  organization={Springer Nature}
}

@article{breitung2021sorm,
  title={SORM, design points, subset simulation, and Markov chain Monte Carlo},
  author={Breitung, Karl},
  journal={ASCE-ASME Journal of Risk and Uncertainty in Engineering Systems, Part A: Civil Engineering},
  volume={7},
  number={4},
  pages={04021052},
  year={2021},
  publisher={American Society of Civil Engineers}
}

@article{nataf1962determination,
  title={Determination des distribution don't les marges sont donnees},
  author={Nataf, Andre},
  journal={Comptes rendus de l'Acad{\'e}mie des Sciences},
  volume={225},
  pages={42--43},
  year={1962}
}

@article{rosenblatt1952remarks,
  title={Remarks on a multivariate transformation},
  author={Rosenblatt, Murray},
  journal={The annals of mathematical statistics},
  volume={23},
  number={3},
  pages={470--472},
  year={1952},
  publisher={JSTOR}
}

@article{li2025relaxed,
  title={Relaxed subset simulation for reliability estimation},
  author={Li, Binbin and Xia, Weili and Liao, Zihan},
  journal={Reliability Engineering \& System Safety},
  volume={264},
  pages={111302},
  year={2025},
  publisher={Elsevier}
}

@incollection{latuszynski2026mcmc,
  title={MCMC Methods for Multi-Modal Distributions},
  author={{\L}atuszy{\'n}ski, Krzysztof and Moores, Matthew T and Stumpf-F{\'e}tizon, Timoth{\'e}e},
  booktitle={Handbook of Markov Chain Monte Carlo},
  pages={368--396},
  year={2026},
  publisher={Chapman and Hall/CRC}
}

@article{pompe2020,
author = {Emilia Pompe and Chris Holmes and Krzysztof Łatuszyński},
title = {{A framework for adaptive MCMC targeting multimodal distributions}},
volume = {48},
journal = {The Annals of Statistics},
number = {5},
publisher = {Institute of Mathematical Statistics},
pages = {2930 -- 2952},
year = {2020},
doi = {10.1214/19-AOS1916},
URL = {https://doi.org/10.1214/19-AOS1916}
}

@article{neal2011mcmc,
  title={MCMC using Hamiltonian dynamics},
  author={Neal, Radford M},
  journal={Handbook of markov chain monte carlo},
  pages={47--95},
  year={2011},
  publisher={Chapman and Hall/CRC}
}

@article{thaler2024reliability,
  title={Reliability analysis of complex systems using subset simulations with Hamiltonian Neural Networks},
  author={Thaler, Denny and Dhulipala, Somayajulu LN and Bamer, Franz and Markert, Bernd and Shields, Michael D},
  journal={Structural Safety},
  volume={109},
  pages={102475},
  year={2024},
  publisher={Elsevier}
}

@inproceedings{ahn2013distributed,
  title={Distributed and adaptive darting Monte Carlo through regenerations},
  author={Ahn, Sungjin and Chen, Yutian and Welling, Max},
  booktitle={Artificial Intelligence and Statistics},
  pages={108--116},
  year={2013},
  organization={PMLR}
}

@article{gabrie2022adaptive,
  title={Adaptive Monte Carlo augmented with normalizing flows},
  author={Gabri{\'e}, Marylou and Rotskoff, Grant M and Vanden-Eijnden, Eric},
  journal={Proceedings of the National Academy of Sciences},
  volume={119},
  number={10},
  pages={e2109420119},
  year={2022},
  publisher={National Academy of Sciences}
}

@inproceedings{Geyer1991,
  author    = {Geyer, Charles J},
  title     = {Markov Chain Monte Carlo Maximum Likelihood},
  booktitle = {Computing Science and Statistics: Proceedings of the 23rd Symposium on the Interface},
  pages     = {156},
  year      = {1991},
  address   = {New York},
  publisher = {American Statistical Association}
}

\end{document}